\documentclass[12pt]{article}
\usepackage{graphicx}
\usepackage{epstopdf}
\usepackage{amssymb}
\usepackage{graphicx,amsfonts,amsmath,bm,epsfig,color,latexsym}
\usepackage{subfig}
\newcommand{\doublespacing}{\let\CS=\@currsize\renewcommand{\baselinesstrech}
	{2.0}\tiny\CS}
\begin{document}
	\newcommand{\bd}{\begin{document}}
		\newcommand{\ed}{\end{document}}
	\newcommand{\bc}{\begin{center}}
		\newcommand{\ec}{\end{center}}
	\newcommand{\bfr}{\begin{flushright}}
		\newcommand{\efr}{\end{flushright}}
	\newcommand{\lt}{\left}
	\newcommand{\rt}{\right}
	\newcommand{\vs}{\vspace}
	\newcommand{\hs}{\hspace}
	\newcommand{\beq}{\begin{equation}}
		\newcommand{\eeq}{\end{equation}}
	\newcommand{\lb}{\linebreak}
	\newcommand{\pb}{\pagebreak}
	\newcommand{\mb}{\makebox}
	\newcommand{\fb}{\framebox}
	\newcommand{\mc}{\multicolumn}
	\newcommand{\ben}{\begin{enumerate}}
		\newcommand{\een}{\end{enumerate}}
	\newcommand{\bit}{\begin{itemize}}
		\newcommand{\eit}{\end{itemize}}
	\newcommand{\oln}{\overline}
	\newcommand{\un}{\underline}
	\newcommand{\lefq}{\lefteqn}
	\newcommand{\ba}{\begin{array}}
		\newcommand{\ea}{\end{array}}
	\newcommand{\beqa}{\begin{eqnarray}}
		\newcommand{\eeqa}{\end{eqnarray}}
	\newcommand{\beqas}{\begin{eqnarray*}}
		\newcommand{\eeqas}{\end{eqnarray*}}
	\newcommand{\bfg}{\begin{figure}}
		\newcommand{\efg}{\end{figure}}
	\newcommand{\bds}{\begin{displaymath}}
		\newcommand{\eds}{\end{displaymath}}
	\newcommand{\btb}{\begin{tabbing}}
		\newcommand{\etb}{\end{tabbing}}
	\newcommand{\para}{\parallel}
	\newcommand{\pad}{\partial}
	\newcommand{\nn}{\nonumber}
	\newcommand{\la}{\leftarrow}
	\newcommand{\ra}{\rightarrow}
	\newcommand{\lgla}{\longleftarrow}
	\newcommand{\lgra}{\longrightarrow}
	\newcommand{\La}{\Leftarrow}\newcommand{\Ra}{\Rightarrow}
	\newcommand{\Lra}{\Leftrightarrow}
	\newcommand{\Lgla}{\Longleftarrow}
	\newcommand{\Lgra}{\Longrightarrow}
	\newcommand{\lan}{\langle}
	\newcommand{\ran}{\rangle}
	\renewcommand{\a}{\alpha}
	\renewcommand{\b}{\beta}
	\newcommand{\g}{\gamma}
	\newcommand{\G}{\Gamma}
	\renewcommand{\d}{\delta}
	\newcommand{\eps}{\epsilon}
	\newcommand{\Th}{\Theta}
	\newcommand{\s}{\sigma}
	\newcommand{\lam}{\lambda}
	\newcommand{\D}{\Delta}
	\newcommand{\ds}{\displaystyle}
	\newcommand{\vare}{E}
	\newcommand{\pr}{\prime}
	\newcommand{\ro}{\rho}
	\newcommand{\nab}{\nabla}
	\newcommand{\m}{\mu}
	\newcommand{\n}{\nu}
	\newcommand{\Sg}{\Sigma}
	\newcommand{\p}{\pi}
	\newcommand{\R}{I\!\!R}
	\newcommand{\om}{\omega}
	\newcommand{\Om}{\Omega}
	\newcommand{\ovra}{\overrightarrow}
	\newcommand{\ze}{\zeta}
	\newcommand{\vart}{\vartheta}
	\newcommand{\tri}{\triangle}
	\newcommand{\f}{\frac}
	\newcommand{\iny}{\infty}
	\newcommand{\pro}{\propto}
	\renewcommand{\arraystretch}{1.25}
	
	\bc {\large\bf Chirped Elliptic Waves: Coupled Helmholtz Equations} \ec
	
	\vs{1cm}

	\bc
	{\it {\bf Naresh Saha}{\footnote{e-mail : naresh\_r@isical.ac.in,}},
		{\bf Barnana Roy}{\footnote{e-mail : taturoy@gmail.com}}\\
		Physics \& Applied Mathematics Unit, \\
		Indian Statistical Institute, \\
		Kolkata - 700 108, India.\\
		 
		{\bf Avinash Khare}{\footnote{e-mail : avinashkhare45@gmail.com}}\\
		Department of Physics,\\
		Sabitribai Phule Pune University,\\
		Pune - 411007, India.}\ec
	\vs{1.2cm}
	
	\bc {\large {\un{Abstract}}} \ec 
	Exact chirped elliptic wave solutions are obtained within the framework of  coupled cubic nonlinear Helmholtz equations in the presence of non-Kerr nonlinearity like self steepening and self frequency shift. It is shown that, for a particular combination of the self steepening and the self frequency shift parameters, the associated nontrivial phase gives rise to chirp reversal across the solitary wave profile. But a different combination of non-Kerr terms leads to chirping but no chirp reversal. The effect of nonparaxial parameter on physical quantities such as intensity, speed and pulse-width of the elliptic waves is studied too. It is found that the speed of the solitary wave can be tuned by altering the nonparaxial parameter. Stable propagation of these nonparaxial elliptic waves is achieved by an appropriate choice of parameters.

%	\newpage
\section{Introduction}
Recently, there has been a lot of interest in the study of nonparaxial beam (pulse) propagation in optical waveguide (fiber) modeled by scalar ( or vector) nonlinear Helmholtz (NLH) equation with Kerr like nonlnearities along with spatial dispersion arising from the breakdown of paraxial approximation also called slowly varying envelope approximation (SVEA). Paraxial approximation holds when the width of optical beam (pulse) of sufficiently low intensity is broader than the carrier wavelength and also when the beam (pulse) propagates along (or at negligible angle) with respect to the reference axis. If the beam (pulse) fails to satisfy at least any one of the aformentioned properties then it is said to be a nonparaxial beam (pulse) \cite{lax,christ20}. When broad beam (pulse) propagate at arbitrary angles with respect to the reference direction, it is referred to as Helmholtz nonparaxiality. The latter has been considered in \cite{cham11,chris56,cham12,cham13} and nonparaxial solitons have been obtained. Also, exact analytical soliton solutions to scalar Helmholtz equation with focusing, defocusing Kerr nonlinearity, power law and polynomial nonlinearity as well as in anomalous and normal dispersion regimes are known \cite{cham11,cham19}. Relativistic and pseudo relativistic formulation of the scalar Helmholtz equation is considered in \cite{chris70,chris700,chris701} and exact analytical soliton solutions in cubic, cubic-quintic and saturable nonlinearity have been obtained.\\
Scalar and vector nonlinear nonparaxial evolution equations 
are developed in \cite{blair} for propagation in two dimensions. Exact and 
approximate solutions to these equations are obtained and are shown to exhibit 
quasi-soliton behavior based on propagation and collision studies. 
The vector extension of the scalar NLH equation to describe multicomponent beam (pulse) evolution in the Kerr type media has been introduced for the first time in \cite{christ1} and is referred to as Helmholtz-Manakov equation. Exact analytical bright-bright and bright-dark vector soliton solutions in self-focusing Kerr media and dark-bright and dark-dark soliton solutions for self-defocusing Kerr-media have been obtained. It can be recalled that in the paraxial regime, the vector extension of the nonlinear Schr\'odinger eqution was proposed by Manakov\cite{mana}. The elliptic and the hyperbolic solitary wave solutions to the coupled NLH system are obtained in \cite{tamil} and effect of non-paraxiality on speed, pulse-width and amplitude of various solutions are explored. Coupled nonparaxial (2+1) dimensional nonlinear Schr\"odinger equation has been considered in \cite{kum} and bright-bright, dark-dark and bright-dark soliton solutions are obtained. Modulation instability of the system of equations is studied too.\\
On the other hand, in recent times, a flurry of research activities \cite{himu}
can be seen on the chirped femtosecond solitons because of their important 
applications in optical pulse fiber amplifiers/compressors and long haul links 
\cite{krug,cun1}. In most of these works non-Kerr terms like self steepening 
and self frequency shift have been added to nonlinear Schr\"odinger equation 
(paraxial regime) as these are important in the study of propagation of 
ultrashort pulses \cite{agra}. It is relevant to mention here a particular 
property of a chirped laser pulse i.e. the laser frequency changes for the 
duration of the pulse: a positively (negatively) chirped pulse implies that the
laser frequency increases (decreases) with time. The positive chirp produces 
optical pule broadening and the negative chirp compensates the broadening by 
narrowing the optical pulse \cite{felix}.  Importantly, the development of high
repetition rate generators of high quality ultrashort pulses play a crucial 
role for the increase of transmission capacity in fiber optic networks, as well
as for optical signal processing \cite{yamma}. In this context, non linear 
transformation of a dual frequency signal in an optical fibre is an attractive 
method for applications to ultra high speed telecommunications. The repetition 
rate of the generated pulses can be tuned by adjusting frequencies of the two 
input waves. For this kind of ultrashort pulse train generation from the 
beating of two mode signals, it is necessary to investigate periodic wave 
solutions of coupled NLSE \cite{chow} (paraxial regime). Also in this regime, 
passively mode locked fibre lasers under certain condition can deliver a 
continuous wave train of picosecond or femtosecond pulses at a repetition rate 
higher than 100 GHz \cite{jochen}, with essentially discrete Fourier spectra of
synchronized modes, which can be effectively described by elliptic waves 
\cite{kartashov}. These results in paraxial domain have motivated us to 
consider chirped elliptic waves in the coupled NLH system (non paraxial regime)
in the presence of self steepening and self frequency shift. The
inclusion of the non-Kerr terms may lead to important physical effects (which 
actually happens as we shall see below), and, therefore, this is a topic well 
worth investigating. So far as we are aware off, these structures in the 
nonparaxial regime of the coupled NLH system have not been reported so far in 
the literature. In the non paraxial regime, the formation and propagation of 
chirped elliptic and solitary waves in the scalar cubic-quintic NLH system 
without non-Kerr terms has been studied in \cite{tamil1}. Modulation 
instability is studied too. The existence and stability of chirped hyperbolic 
solitary waves for cubic coupled NLH equation in the presence of self 
steepening and self frequency shift are studied in \cite{naresh}. The novel 
physical effect that emerges from the inclusion of non-Kerr terms is that the 
presence or absence of chirp reversal depends on the relation between the self 
steepening and self frequency shift parameters. 
Thus the main aim of this article is to obtain analytically chirped elliptic 
wave solutions to coupled NLH equation with non-Kerr nonlinear terms like 
self-steepening and self frequency shift. We have been able to find three 
families of chirped elliptic waves together with the conditions that specify 
the parameters.  As obtained in the case of the chirped hyperbolic waves 
\cite{naresh}, it has been found that for one particular combination of the 
self steepening and self frequency shift parameters, the associated phase has 
two intensity dependent terms. One varies as the reciprocal of the intensity 
while the other which depends on non-Kerr nonlinearities, is directly 
proportional to the intensity. As a result chirp reversal occurs across the 
wave profile. For a different relation between the non-Kerr terms, there exists
only one intensity dependent term which is inversely proportional to the 
intensity, resulting in chirping but no chirp reversal. The interesting 
qualitative features of the chirping of the two components associated with the 
effect of nonparaxiality and non-Kerr nonlinearity have been revealed.

\section{Theoretical Model}
The coupled Helmholtz equation in the presence of non-Kerr nonlinear terms describing the propagation of femtosecond pulses in optical fibres without slowly varying envelope approximation is given by 
\begin{equation}\label{1a}
\begin{array}{lcl}
i q_{1z} + \kappa q_{1zz} + \frac{1}{2} q_{1tt}+(\bar{\sigma}_1 |q_1|^2 + \bar{\sigma}_2 |q_2|^2) q_1+\\
i[a_4(|q_1|^2q_1)_t+a_5q_1|q_1|^2_{t}] = 0,
\end{array}
\end{equation}
\begin{equation}\label{2a}
\begin{array}{lcl}
i q_{2z} + \kappa q_{2zz} + \frac{1}{2} q_{2tt}+(\bar{\sigma}_1 |q_1|^2 + \bar{\sigma}_2 |q_2|^2) q_2+\\
i[a_4(|q_2|^2q_2)_t+a_5q_2|q_2|^2_{t}]  = 0,
\end{array}
\end{equation}
where the subscripts $z$ and $t$ represent the longitudinal and transverse coordinates respectively, $q_1$ and $q_2$ are the envelopes of the first and second component respectively. The second term represent spatial dispersion originating from nonparaxial effect with $\kappa$ being the non paraxial parameter and the third term represent group velocity dispersion while $\bar{\sigma}_i, i = 1,2$ are the nonlinearity coefficients. For $\bar{\sigma_1} = 
\bar{\sigma_2} = \pm 1$, the above equations reduce to Helmholtz-Manakov 
system introduced in \cite{christ1}. The term proportional to $a_4$ represents self steepening \cite{and} which produces a temporal pulse
distortion leading to the development of an optical shock on the trailing edge of the pulse unless
balanced by the dispersion \cite{oliv}. This phenomenon is due to the intensity dependence of the group
velocity that makes the peak of the pulse move slower than the peak \cite{shar}. The last term proportional
to $a_5$ has its origin in the delayed Raman response \cite{shar} which forces the pulse to undergo a frequency
shift known as self frequency shift \cite{mit}.

To construct chirped elliptic wave solutions of 
Eqns. (\ref{1a}) and (\ref{2a}), we start with the following traveling wave ansatz
\begin{equation}\label{3}
q_j(z,t) = f_j(\xi) e^{i[\phi_j(\xi)-k_j z +\zeta_j]}\,,~~j = 1, 2,
\end{equation}
where $\xi = \beta(t-vz)$, $v = \frac{1}{u}$, $u$ is the group velocity of 
the wave packet, $k_j, \zeta_j, j = 1,2$ are the wave numbers and real constants respectively. 
Substituting the ansatz (\ref{3}) into Eqns. (\ref{1a}) and (\ref{2a}), 
collecting the imaginary parts and integrating the resulting equations, we 
get
\begin{eqnarray}
&&\phi_1^{\prime}(\xi) = \frac{c_1}{f_1^2} + \frac{a_1}{2a}
-\frac{\beta(3a_4+2a_5)f_1^2}{4a}\label{4},\\
&&\phi_2^{\prime}(\xi) = \frac{c_2}{f_2^2} + \frac{a_2}{2a}
-\frac{\beta(3a_4+2a_5)f_2^2}{4a}\label{4a}\,,
\end{eqnarray}
where $c_1$ and $c_2$ are the integration constants while $a, a_1, a_2$ are 
given by 
\begin{eqnarray}\label{5}
&&a = \beta^2(\kappa v^2 + 1/2) > 0\,,~~a_1 = \beta v (1-2\kappa k_1)\,, \nonumber \\
&&a_2 = \beta v(1-2\kappa k_2).
\end{eqnarray}
Hence the chirping comes out to be
\begin{equation}\label{6}
\delta \omega_i = - \beta \left(\frac{a_i}{2a} + \frac{c_i}{f_i^2}
-\frac{\beta(3a_4+2a_5)f_i^2}{4a}\right),~~
i = 1,2\,,
\end{equation}
which depends on velocity $u$, wave numbers $k_i, i = 1,2$, nonparaxial 
parameter $\kappa$ and the intensities $f^{2}_{i}, i = 1,2$, of the chirped 
pulse. 
It is worth noticing that the first term on the right hand side of Eqn. (\ref{6}) is constant. The second term is of kinematic origin and is inversely proportional to the 
intensity while the last term is directly proportional to the intensity so that it leads 
to chirping that is inverse to that coming from the second
term. 
On substituting the ansatz (\ref{3}) into Eqns. (\ref{1a}) 
and (\ref{2a}), collecting the real parts and utilizing Eqns. (\ref{4}), (\ref{4a}) we 
get the following equations
\begin{eqnarray}
&&f_{1}^{''} + d_1 f_1 +(\sigma_1 f_{1}^{2} +\sigma_{2} f_{2}^{2}) f_1+\beta_1f_1^5-\delta_1f_1^3
= \frac{c_{1}^{2}}{f_{1}^{3}},~~~~~\label{7}\\
&&f_{2}^{''} + d_2 f_2 +(\sigma_1 f_{1}^{2} +\sigma_{2} f_{2}^{2}) f_2+\beta_1f_2^5-\delta_2f_2^3 
= \frac{c_{2}^{2}}{f_{2}^{3}},~~~~~\label{8}\,
\end{eqnarray}
where
\begin{eqnarray}\label{9}
&&d_1 = \frac{a_{1}^{2}+4a b_1+2\beta c_1 a(a_4+2a_5)}{4a^2},\, \nonumber \\
&&d_2 = \frac{a_{2}^{2}+4a b_2+2\beta c_2 a(a_4+2a_5)}{4a^2},\, \nonumber \\
&&\sigma_{1,2} = \frac{\bar{\sigma}_{1,2}}{a}\,,~~\delta_1=\frac{\beta a_4 a_1}
{2 a^2},\, \\
&&\delta_2=\frac{\beta a_4 a_2}{2 a^2}\,,~~b_1 = k_1(1-\kappa k_1)\,,~~b_2 
= k_2(1-\kappa k_2)\,, \nonumber \\
&&\beta_1=\frac{\beta^2(3a_4+2a_5)(a_4-2a_5)}{16a^2}.\, \nonumber
\end{eqnarray} 
We shall obtain analytical solutions of the 
coupled Eqns. (\ref{7}) and (\ref{8}) in case $f_2$ and $f_1$ are 
proportional to each other, i.e. 
$f_2(\xi) = \alpha f_1(\xi)$, with $\alpha$ being a real 
number. For this we have 
\begin{equation}\label{10}
\beta_1 = 0\,,~~d_1 = d_2\,,~~c_{2}^{2} = \alpha^4 c_{1}^{2},\,~~
\delta_1 = \alpha^2 \delta_2\,.
\end{equation}
Now $\beta_1 = 0$ can be satisfied if either
(i) $a_4 = 2a_5\implies$ chirp reversal or 
(ii) $3a_4 +2a_5=0 \implies$ chirping only but no chirp reversal.
We shall focus on the case $a_4 = 2a_5$. Eqn. (\ref{10}) gives
\begin{equation}\label{333}
\alpha^2=\frac{1-2\kappa k_1}{1-2\kappa k_2},
\end{equation}
and
\begin{eqnarray}\label{334}
&&c_1=\frac{\beta (k_2-k_1)(1-2\kappa k_2)}{4a a_4}\,,\rm{when} ~~c_2 = -\alpha^2 c_1\,, \\
&&c_1=\frac{\beta [\kappa(k_1+k_2)-1](1-2\kappa k_2)}{4a a_4\kappa},~\rm{when}~c_2 = \alpha^2 c_1.\, \nonumber
\end{eqnarray}
In what follows, 
we shall take 
$c_1=\frac{\beta (k_2-k_1)(1-2\kappa k_2)}{4aa_4}$, 
$\bar{\sigma_1} = -1$ and $\bar{\sigma_2} = -1$, $\zeta_1=0.1,~\zeta_2=0.2$. 
To solve Eqn. (\ref{7}) we take the ansatz
\begin{equation}\label{10a}
f_1(\xi) = \sqrt{\mu(\xi)}\,,
\end{equation}
so that $\mu$ satisfies the equation
\begin{equation}\label{11}
\frac{d^2 \mu}{d \xi^2} + 4d_1 \mu + 3 \eta_1 \mu^2 
= 2 c_3\,,
\end{equation}
where $c_3$ is an integration constant and $\eta_1=\sigma_1 
+\alpha^2 \sigma_2 - \delta_1$ . At this point it is imperative to test the stability of the results relative to the main assumption $f_2 = \alpha f_1$ i.e. to see whether the solutions obtained hereafter remain stable if there is a slight shift of this proportionality given by
\begin{equation}\label{1s} 
f_2 = \alpha f_1 + \epsilon  g(\xi),
\end{equation}
where $g$ is a function of $\xi$ and $\epsilon$ is a small quantity.
Substitution of Eqn.(\ref{1s}) in Eqns.(\ref{7})  and (\ref{8}) and retaining terms upto $O(\epsilon)$ yields
\begin{equation}\label{2s}
	2\alpha \sigma_2 f_1^2 g(\xi) = 0,
\end{equation}
and
\begin{equation}\label{3s}
	g''(\xi) + [d_2 +\sigma_1 f_1^2 + 3(\sigma_2 - \delta_2) \alpha^2 f_1^2 + 5\beta_1 \alpha^4 f_1^4] g(\xi) = 0.
\end{equation}
The coupled equations (\ref{2s}) and (\ref{3s}) are only satisfied if $g = 0$ thereby showing that our solutions are stable against small perturbations to the $O(\epsilon)$.

\section{Chirped elliptic waves}
 It is important to mention here that for all the solutions that we present here, constraints as given by Eqn. (\ref{10}) must be satisfied. Chirp reversal is shown for one of the components as for the other component it is similar. To check the stability of the nonlinear waves, numerical simulation of the intensity profiles are done.\\
{\bf Solution I.} \\
\begin{equation}\label{512}
\mu(\xi) = A^2 \frac{D + dn(\xi, m)}{E + dn(\xi, m)}\,,
\end{equation}
is an exact solution to Eqn. (\ref{11}) provided either $E = 1$ or 
$E = \sqrt{1-m}$.\\ 
{\bf Case (a). $E = 1$}\\
Eqn. (\ref{512}) is an exact solution provided 
\begin{eqnarray}\label{513}
&&d_1 =\frac{4-5m - D(4+m)}{4 (1-D)},~~~\eta_1 A^2 
= \frac{m}{(1-D)},\, \nonumber \\
&&\frac{2c_3}{A^2} = \frac{2(1-D^2) - m(D+2)}{1-D}.
\end{eqnarray}
The intensity profiles of both the components as a function of $\xi$ 
are given in Fig. 1(a). The plots of the kinematic, higher order and the
combined chirping of $q_1$ as a function of $\xi$ are shown in Fig. 1(b). This plot
clearly shows the chirping reversal. 
Fig.1(c) shows that, chirping of the first component decreases while that of the second component increases as $\kappa$ increases. To start with, similar behaviour is seen (in Fig.1(d)) with respect to $a_5$ as well, but the chirping of both the components saturate as $a_5$ increases further. While the first component shows positive chirping, second component shows negative chirping with respect to both $\kappa$ and $a_5$. Fig.1(e) shows that to begin with, the intensities of both the components initially increase with increasing $\kappa$ but starts 
decreasing after $\kappa$ reaches a certain value. In Fig. 1(f) we show the
behavior of $|v|$ as well as $1/\beta$ as a function of $\kappa$. It is
clearly seen from the figure that while the speed decreases as $\kappa$
increases, the pulse-width $1/\beta$ increases as $\kappa$ increases.
So the speed of the pulse can be decelerated by adjusting $\kappa$. 
Simulation of the intensity profiles of both the components as depicted in Figs.1(g),(h) exhibit stable evolution and pulse compression.\\
 \begin{figure}
 	\centering
 	\subfloat[\label{}]{\includegraphics[width=4.0cm,height=3.0cm]{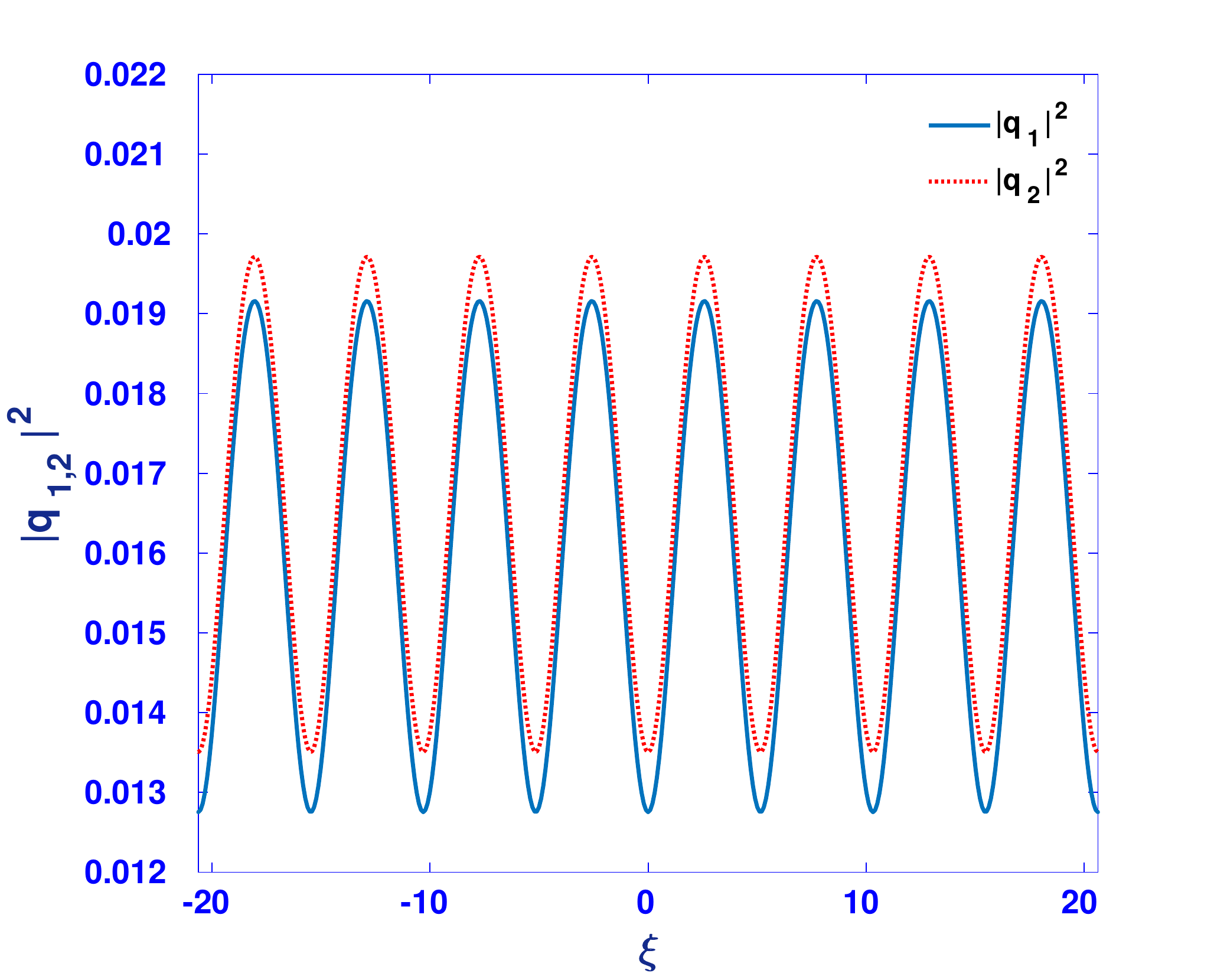}}
 	~
 	\subfloat[\label{}]{\includegraphics[width=4.0cm,height=3.0cm]{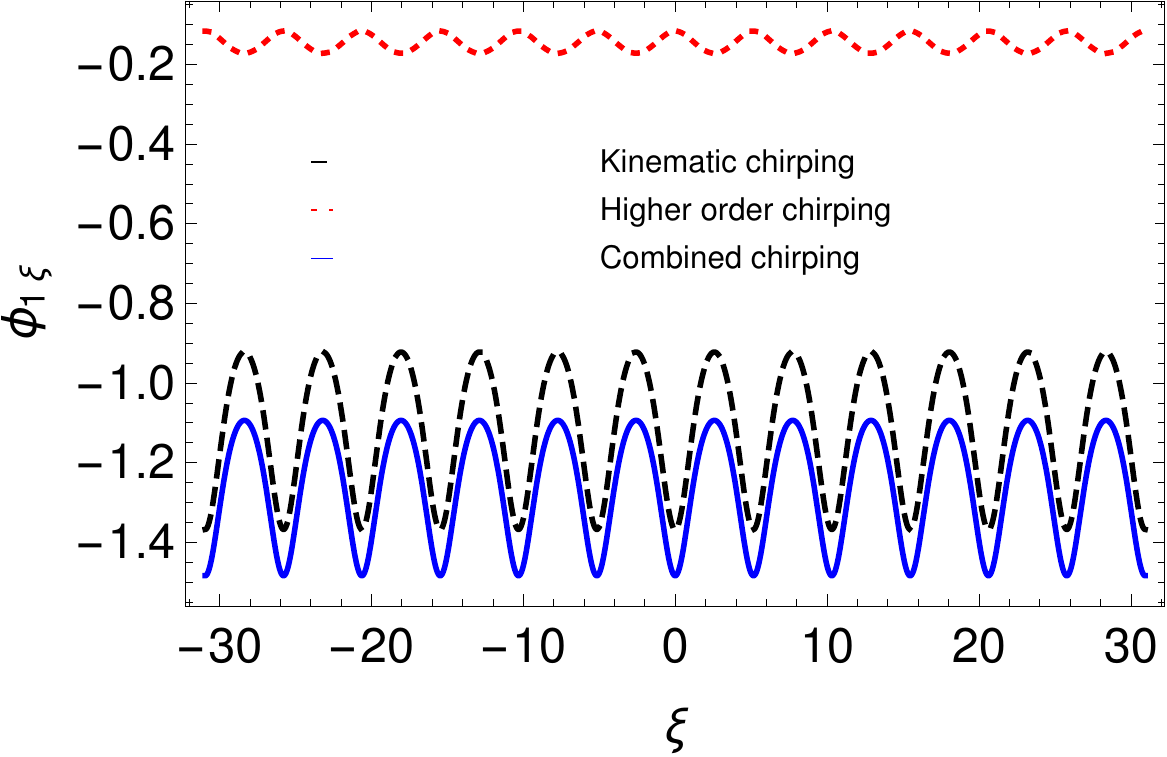}}
 	~
 	\subfloat[\label{}]{\includegraphics[width=4.0cm,height=3.0cm]{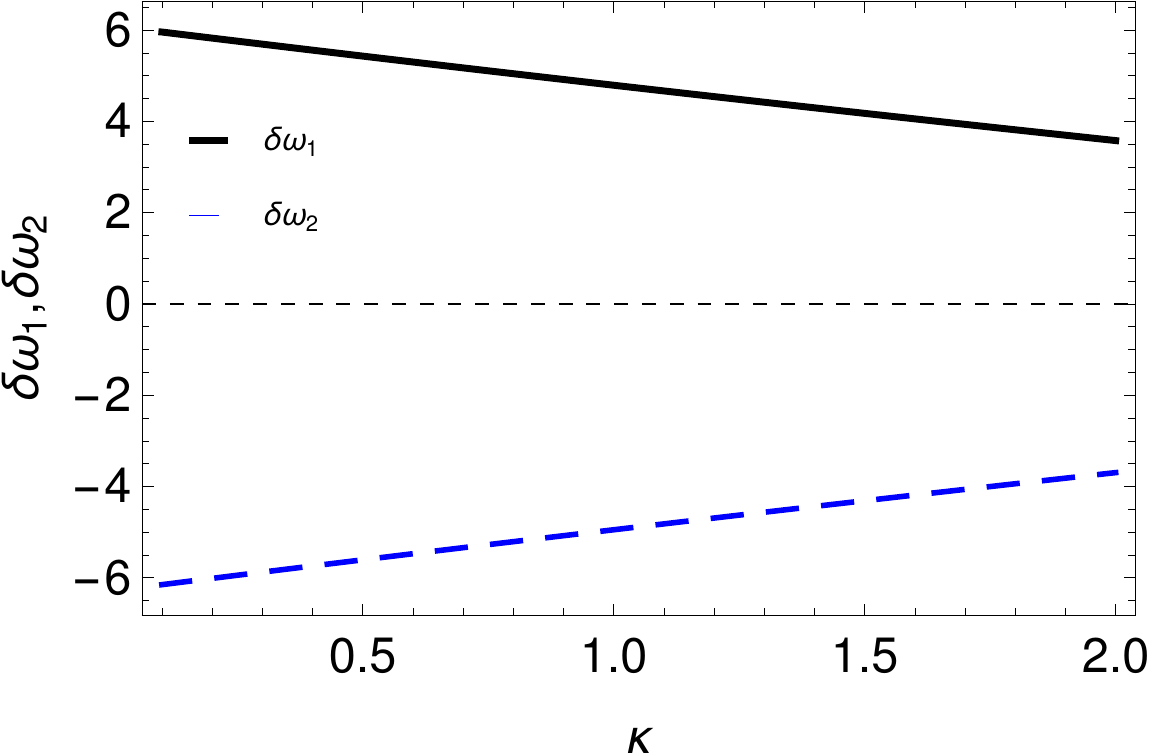}}
 	~
 	\subfloat[\label{}]{\includegraphics[width=4.0cm,height=3.0cm]{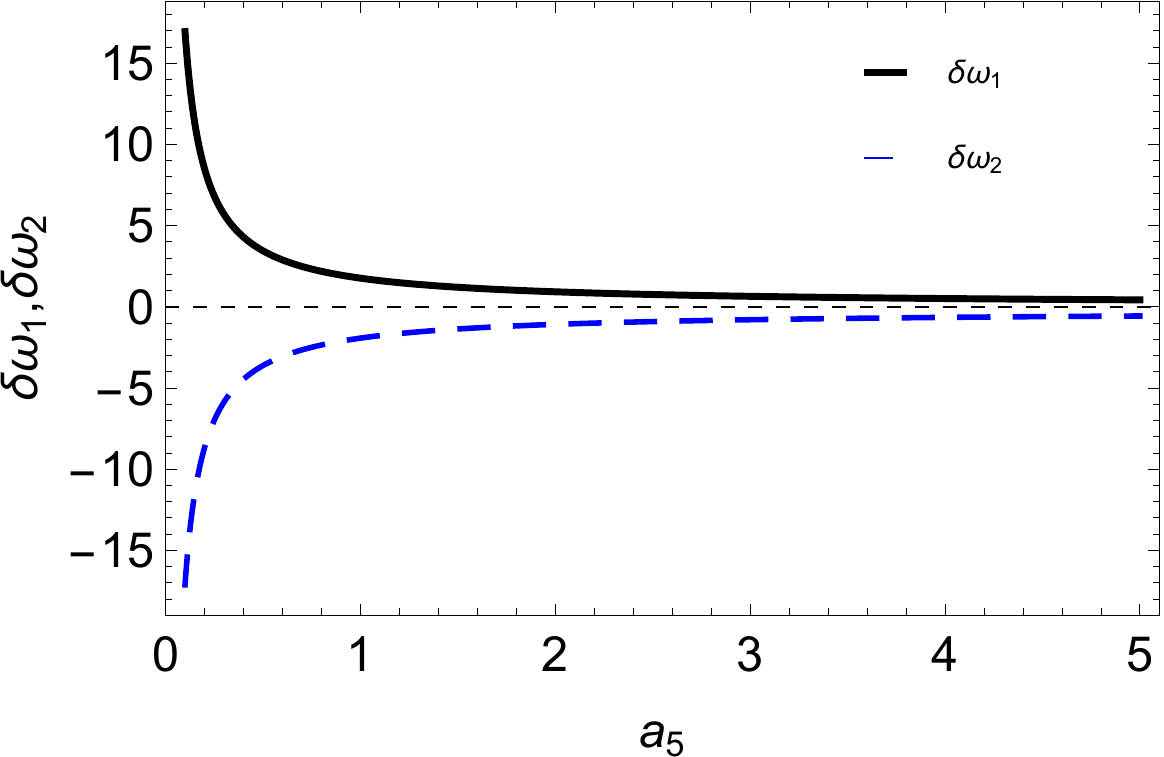}}\\
 	\subfloat[\label{}]{\includegraphics[width=4.0cm,height=3.0cm]{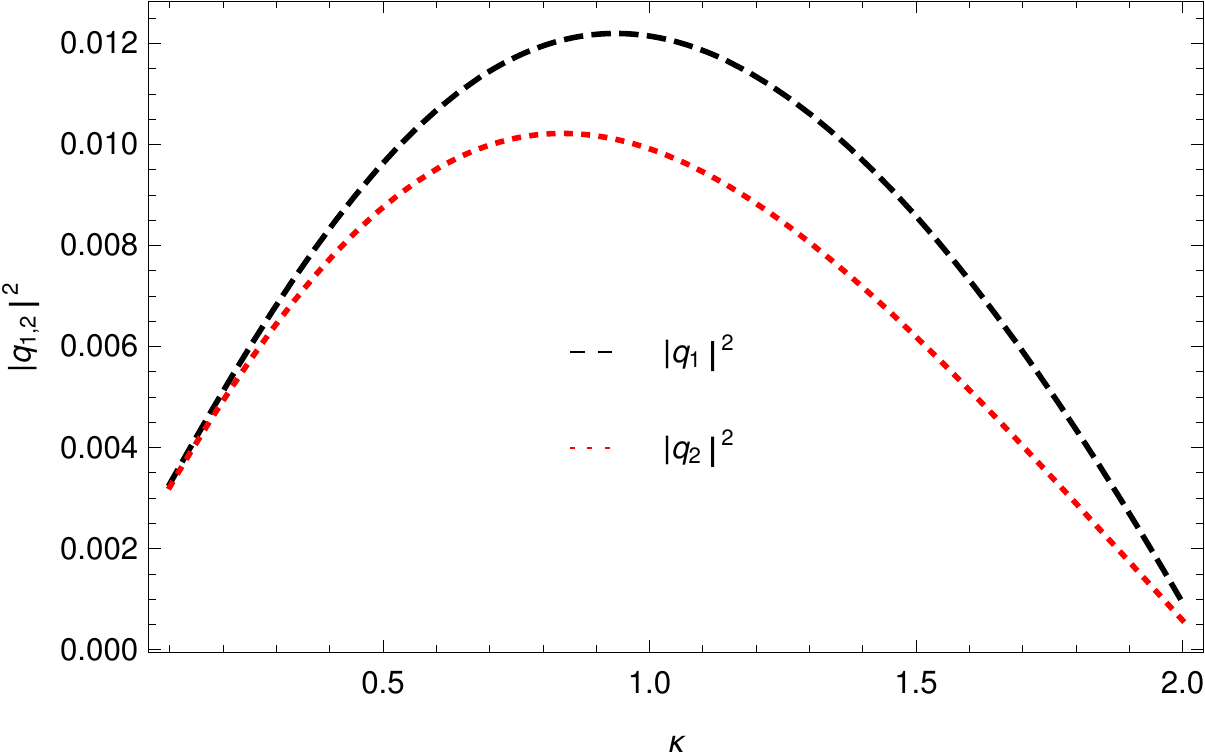}}
 	~~
 	\subfloat[\label{}]{\includegraphics[width=4.0cm,height=3.0cm]{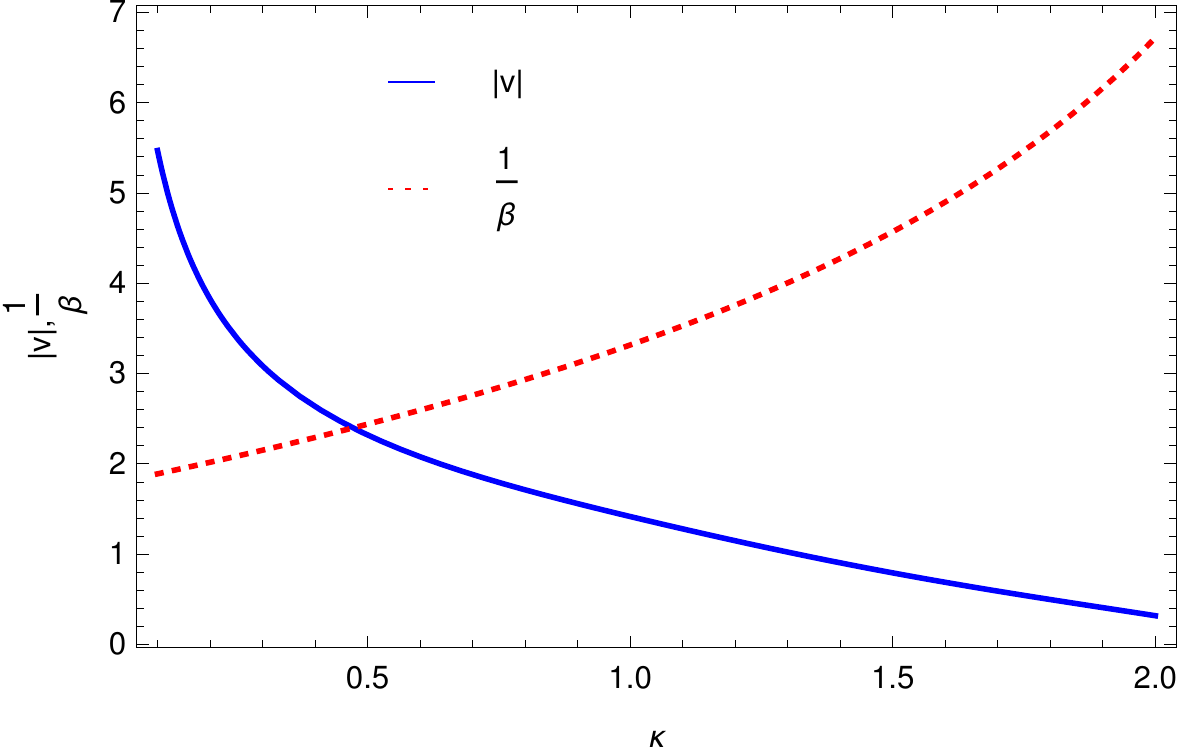}} 
 	~~
 	\subfloat[\label{}]{\includegraphics[width=4.0cm,height=3.0cm]{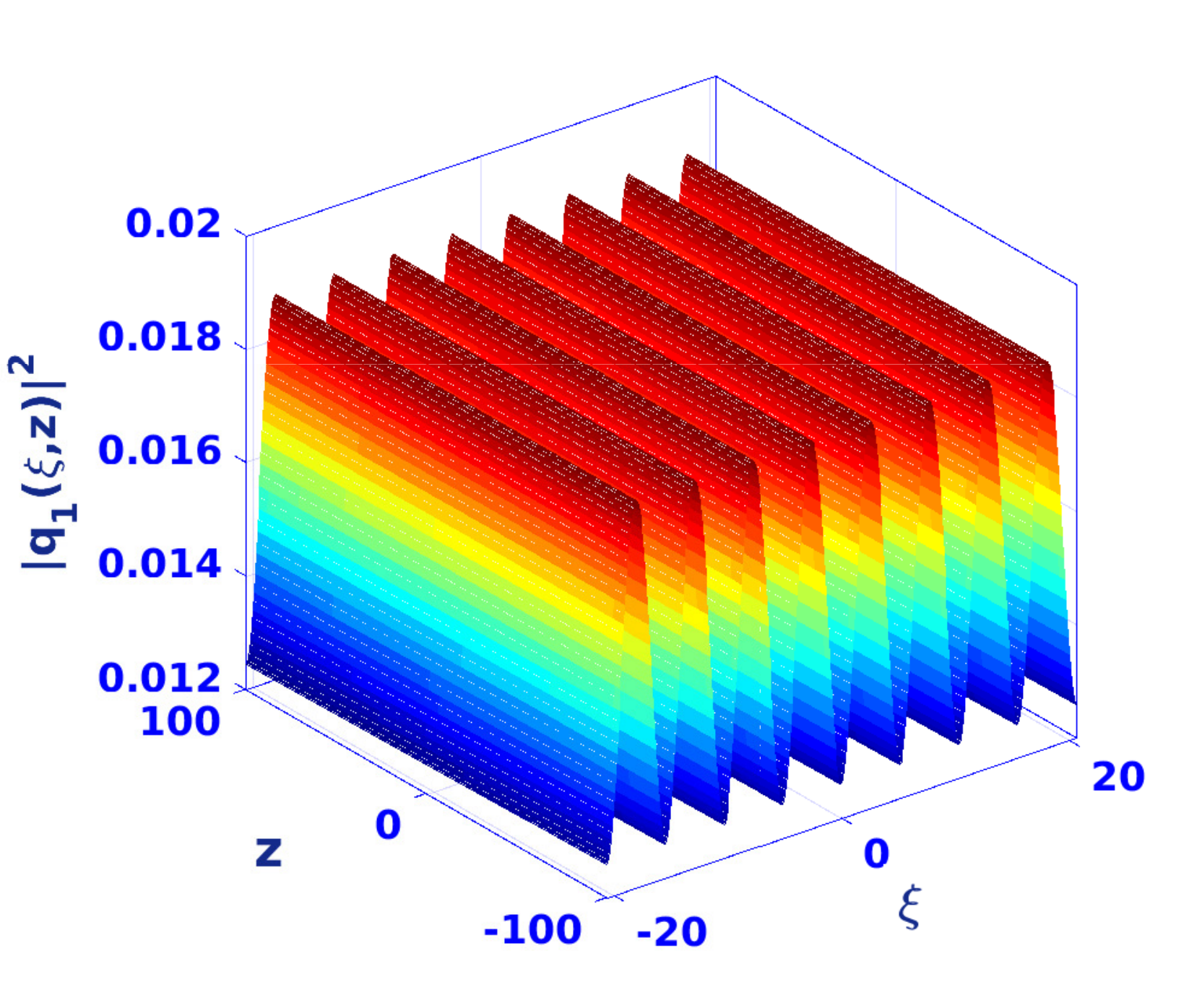}}
 	~~
 	\subfloat[\label{}]{\includegraphics[width=4.0cm,height=3.0cm]{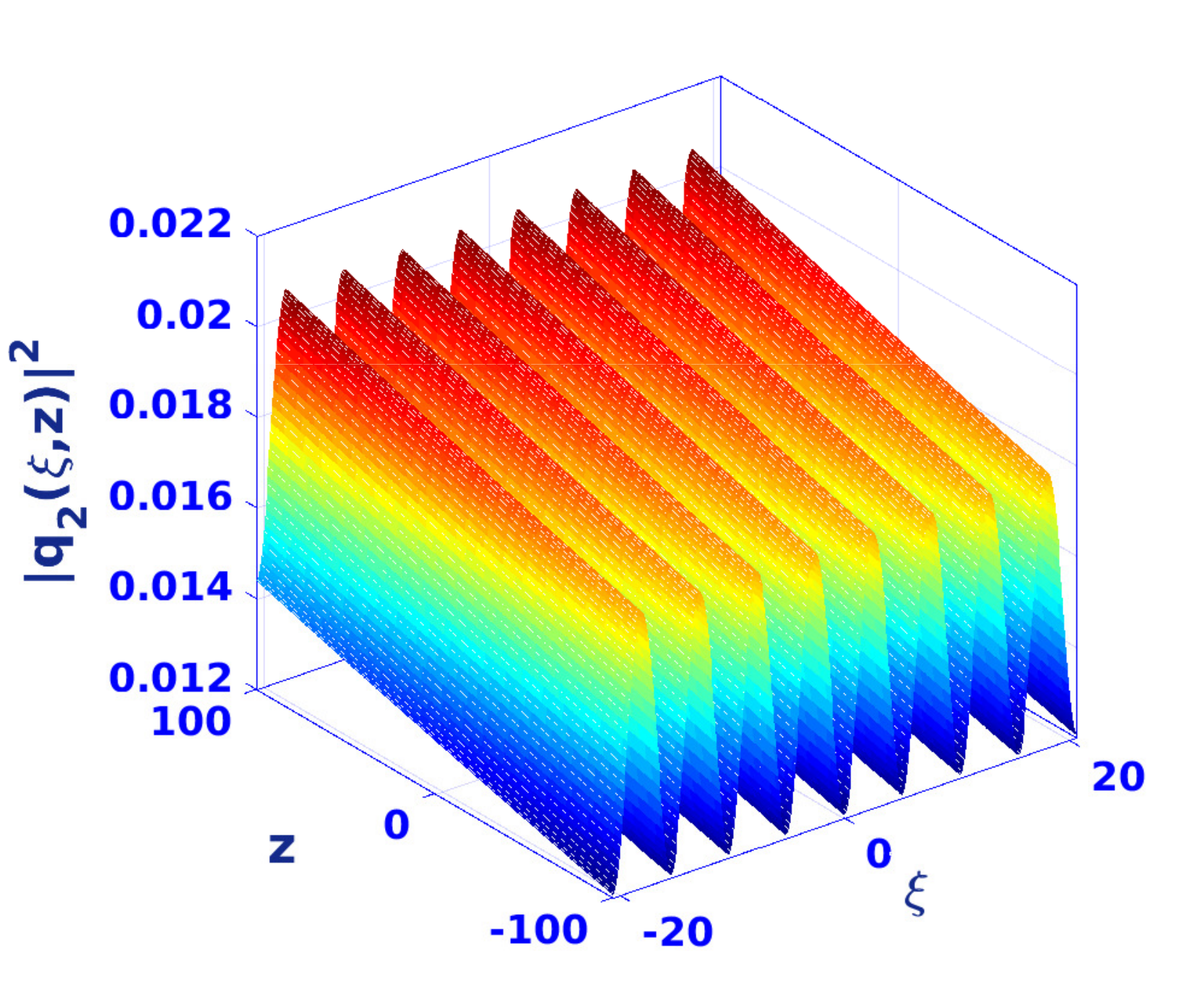}}
 	\caption{For Solution I(Case (a)) for $m = 0.7$ (a) Plot of $|q_1|^2~\rm(blue~solid~line)$,~$|q_2|^2~\rm(red~dotted~line)$ versus $\xi$ for  $\beta=0.2,~\kappa=0.5,~v=1.5,~k_1=0.05,~k_2=0.01,~a_5=0.5,~c_1=-0.0305,~c_2=0.0292,c_3=0.0207,~D=28.005$ at $z=5$, (b) Plot of kinematic (black dashed line), higher order (red dotted line) and combined chirping (blue solid line) of $q_1$ versus $\xi$  for  $\kappa=0.9,~\beta=0.25,~a_5=0.5,~v=0.5,~k_1=0.55,~k_2=0.45,~c_1=-0.0291188,~c_2=-0.00153257,~c_3=0.043065,~D=22.7993$, (c),(d)  Plots of  $\delta \omega_{1,2}$ versus $\kappa$ for $a_5=0.3$, $\delta \omega_{1,2}$ versus $a_5$ for $\kappa=0.3$ respectively and other parameters are $\beta=0.1,~v=0.1,~k_1=0.1,~k_2=0.09,~\xi=5$, 
	(e) Plot of  $|q_1|^2~\rm(black~dashed~line)$,~$|q_2|^2~\rm(red~dotted~line)$ versus $\kappa$ for  $\beta=0.1,~v=2.5,~k_1=0.1,~k_2=0.01,~a_5=0.05,~\xi=5$, (f) Plot of $|v|$ (Blue solid line) versus $\kappa$ for  $\beta=0.1$ and $\frac{1}{\beta}$ (Red dotted line) versus $\kappa$ for $v=0.5$ and other parameters are $k_1=0.5,~k_2=0.3,~D=1.5$, (g),(h) Simulation of the intensity profiles for parameter values same as in (a).} 	
 \end{figure}
Before moving on let us mention that for all the chirped elliptic waves presented hereafter, the behavior of $|v|$ as well as pulse width $1/\beta$ with respect to $\kappa$ is similar to that of Fig. 1(f). So we refrain ourselves from presenting similar plot for other solutions.\\
{\bf Case (b). $E = \sqrt{1-m}$}\\ 
In this case the following parametric conditions are to be satisfied
\begin{eqnarray}\label{21}
&&d_1=\frac{(5m-4)D+(4+m)\sqrt{1-m}}{4(\sqrt{1-m}-D)},\nonumber\\
&&\eta_1 A^2=-\frac{m\sqrt{1-m}}{\sqrt{1-m
	}-D}\,,
 \\
&&\frac{2c_3}{A^2}=\frac{m D + 2(1- D^2)\sqrt{1-m}}{\sqrt{1-m} - D}.\nonumber
\end{eqnarray}
Plots of $q_1^2$ and $q_2^2$ with respect to $\xi$ are given in Figs. 2(a), and the associated chirp reversal for the component $q_2$ is shown Fig.2(b). Fig.2(c) shows that the chirping of the first component initially increases with increasing $\kappa$ but starts decreasing when a certain value of $\kappa$ is reached but that of the second component increases as $\kappa$ increases. 
Chirping of the first component initially decreases with $a_5$ but starts increasing when a certain value of $a_5$ is reached but that of the second component increases as $a_5$ increases as shown in Fig.2(d). While the first component shows positive chirping, second component shows negative chirping when $\kappa$ and $a_5$ increases.
Fig. 2(e) clearly shows that the intensity of both the components decrease as 
$\kappa$ increases. Stable evolution of the intensity profiles with pulse compression are shown in Figs.2(f), (g).
\begin{figure}
	\centering
	\subfloat[\label{}]{\includegraphics[width=4.0cm,height=3.0cm]{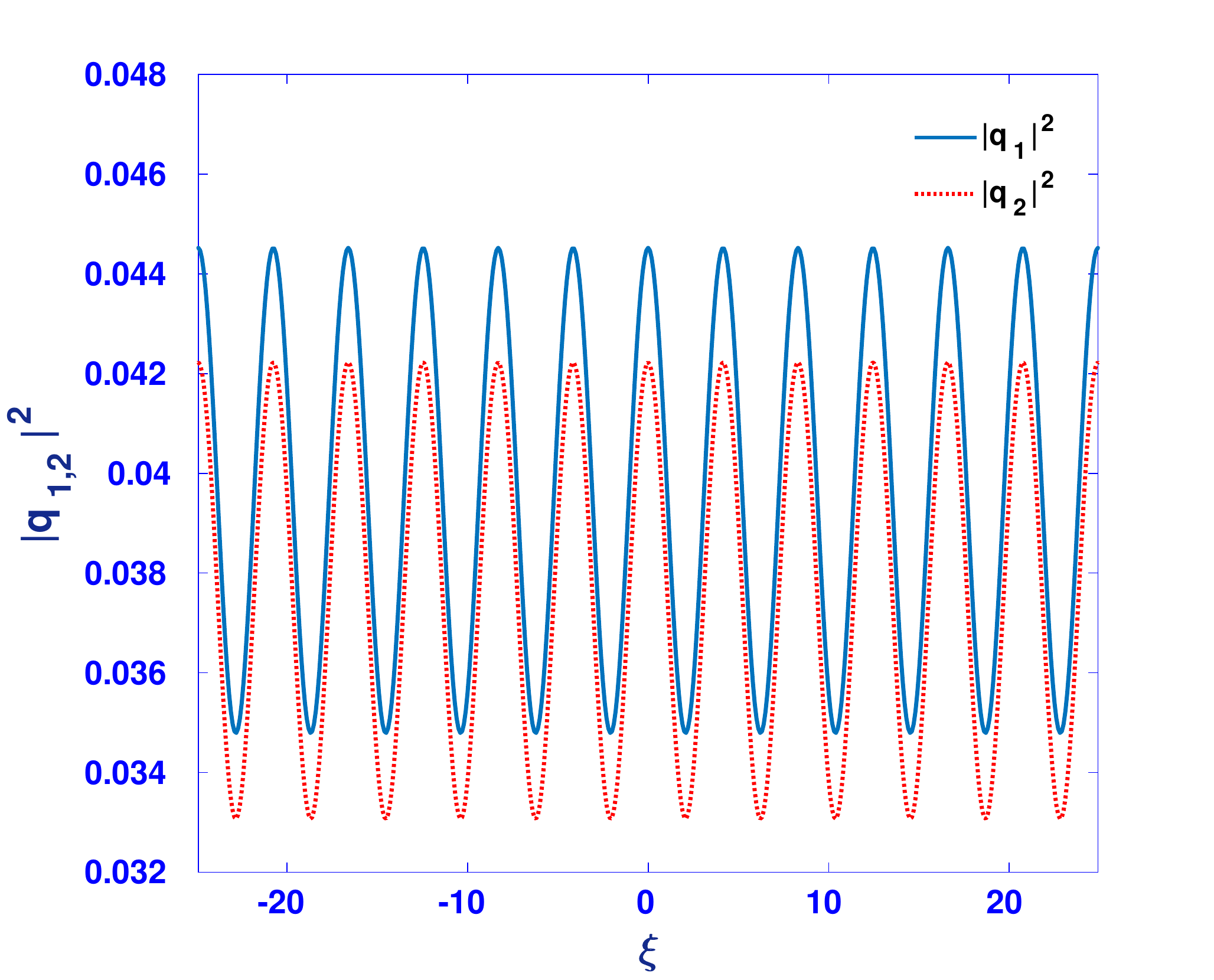}}
	~
	\subfloat[\label{}]{\includegraphics[width=4.0cm,height=3.0cm]{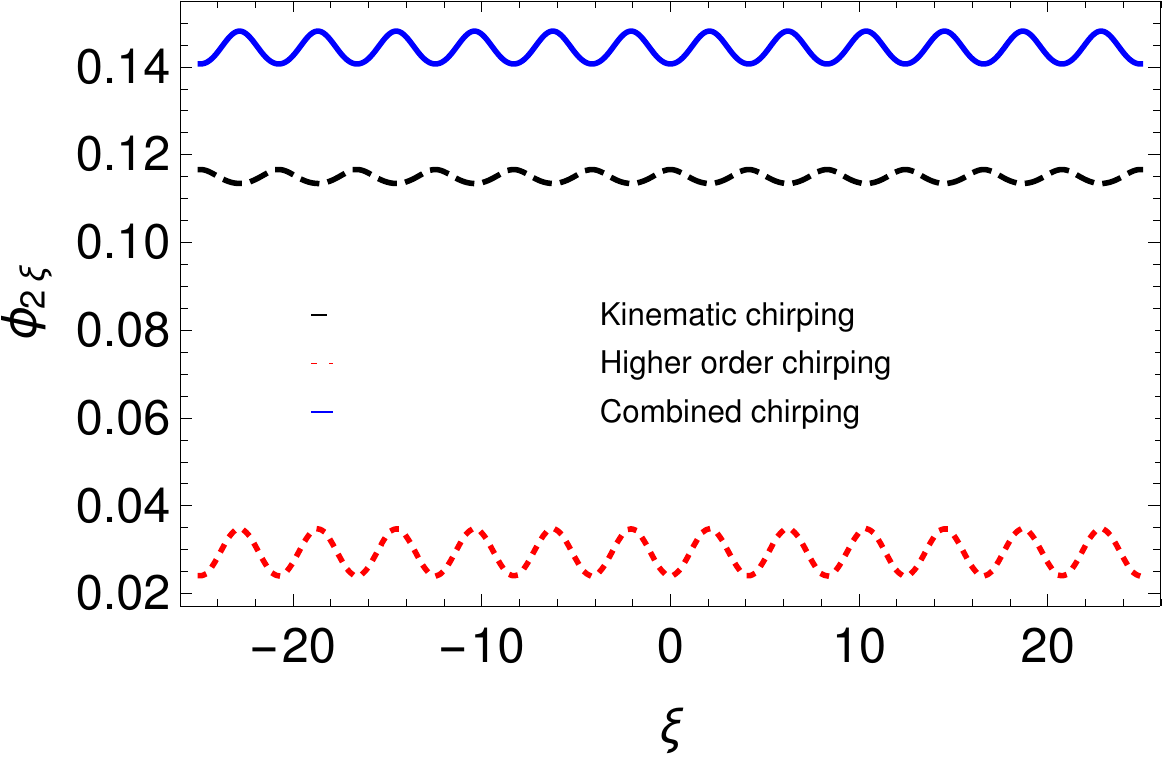}}
	~
	\subfloat[\label{}]{\includegraphics[width=4.0cm,height=3.0cm]{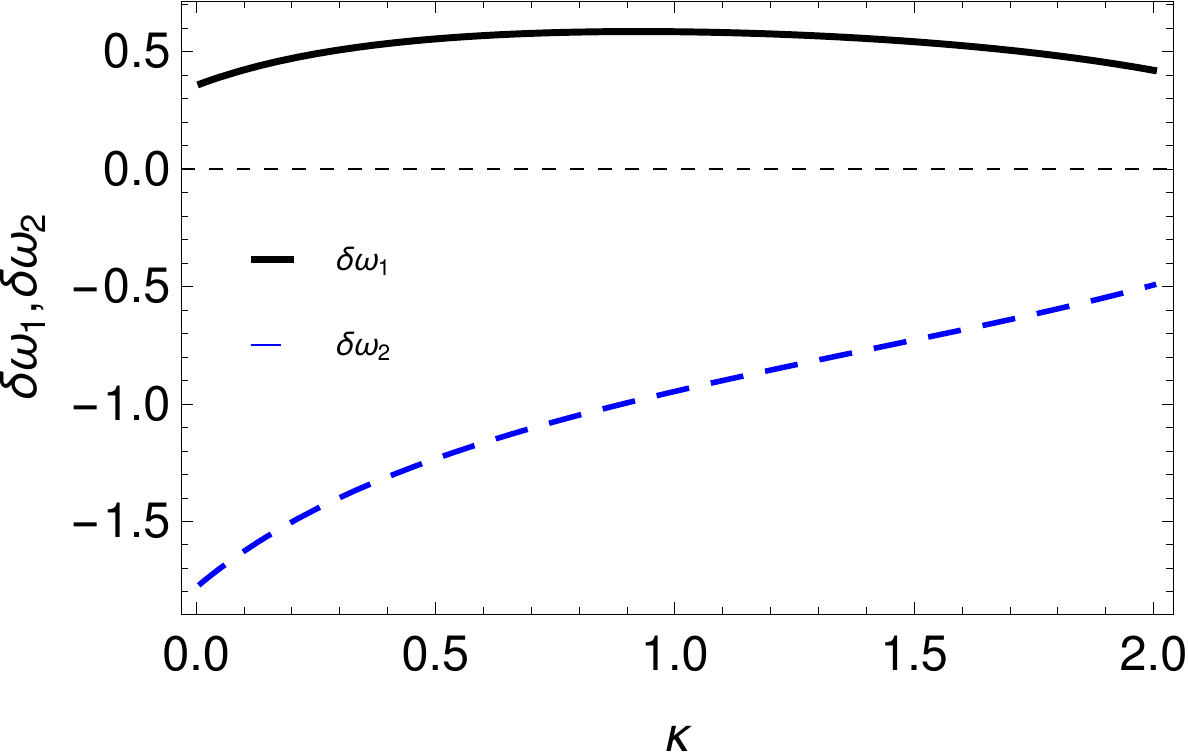}}
	~
	\subfloat[\label{}]{\includegraphics[width=4.0cm,height=3.0cm]{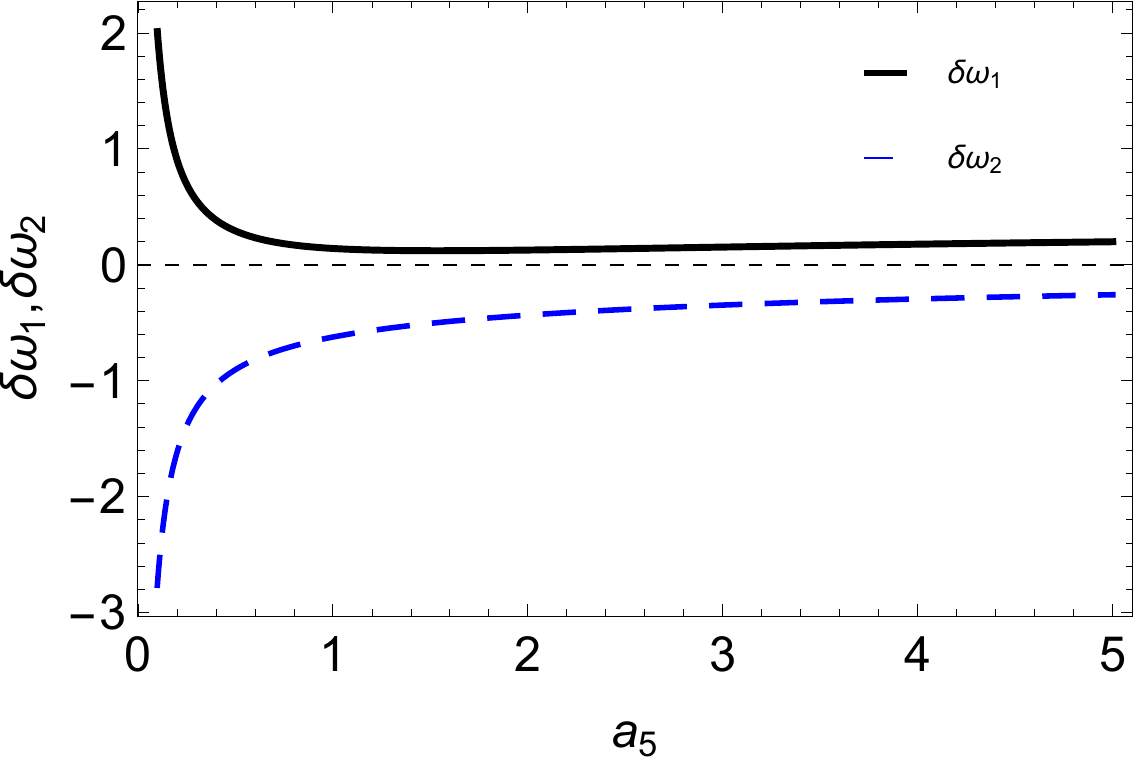}}\\
	\subfloat[\label{}]{\includegraphics[width=4.0cm,height=3.0cm]{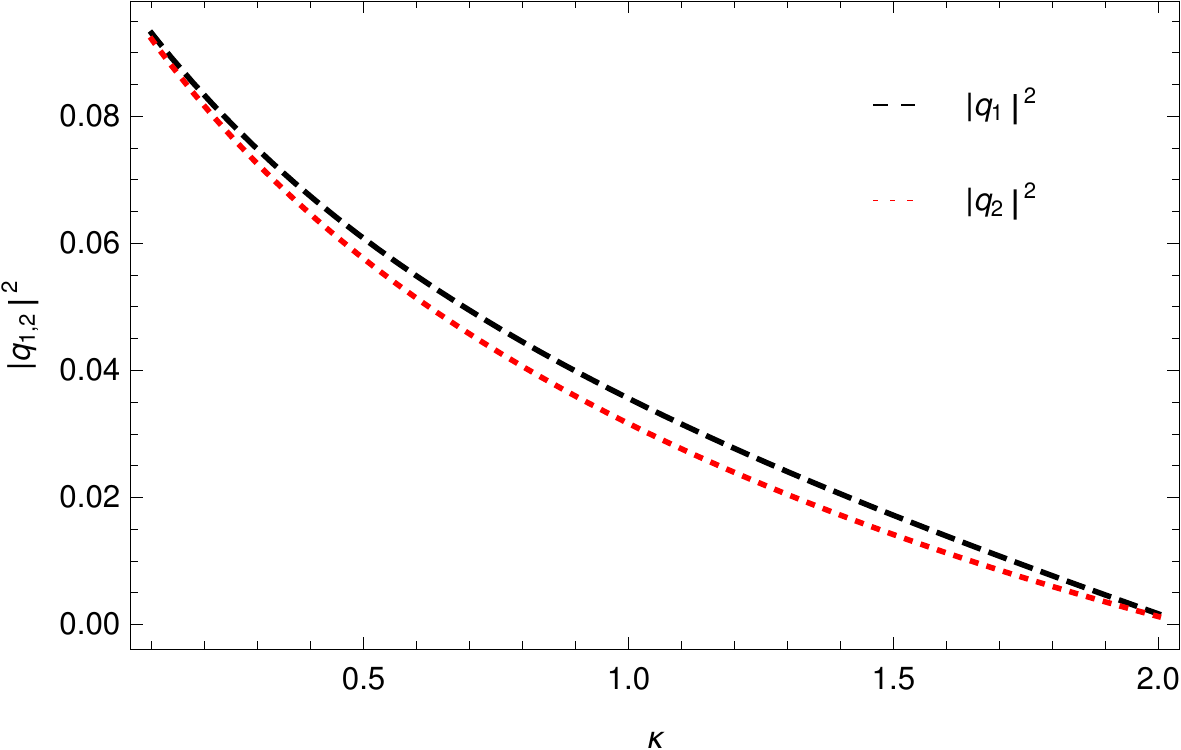}}
	~~
	\subfloat[\label{}]{\includegraphics[width=4.0cm,height=3.0cm]{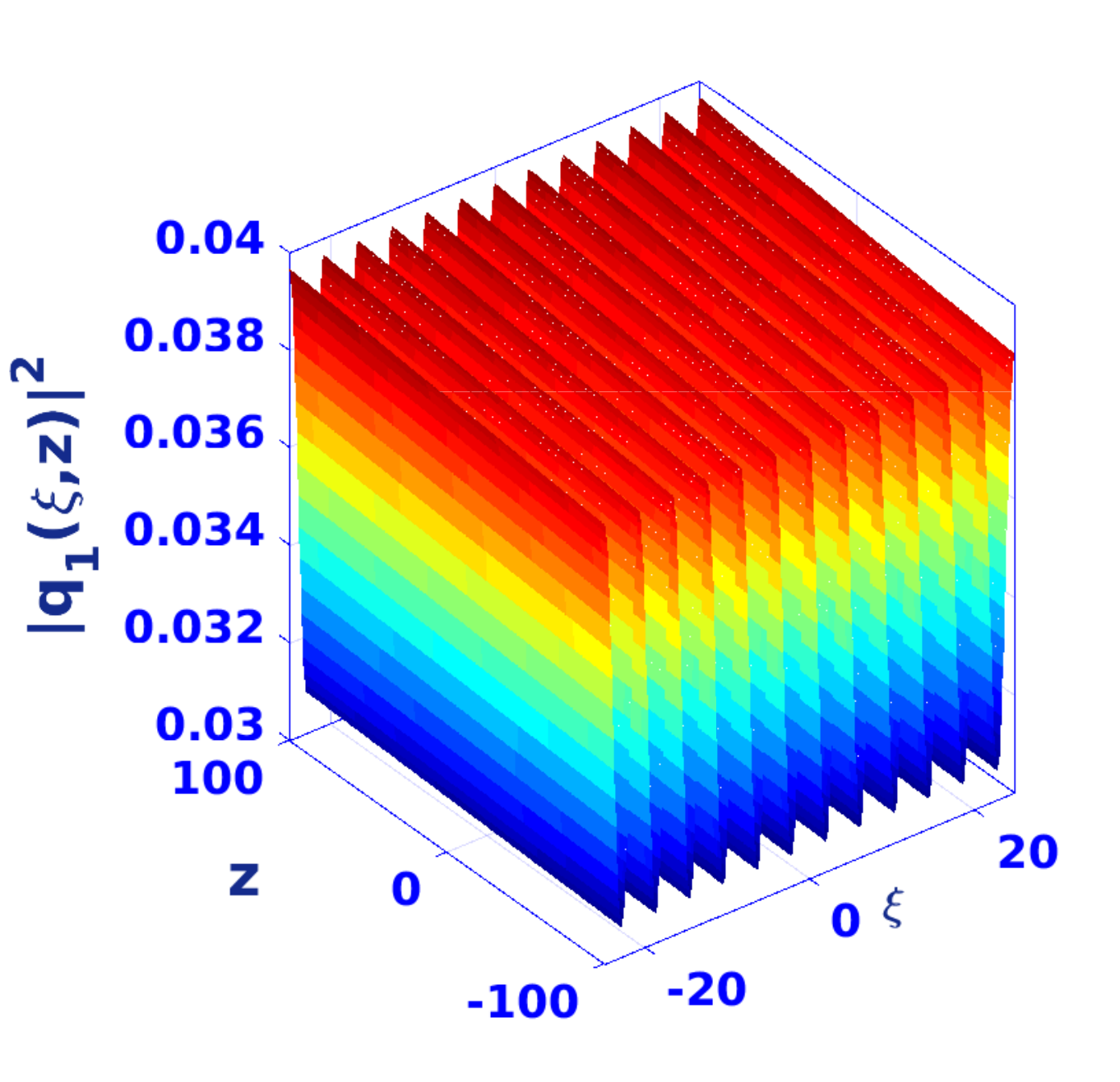}}
	~~
	\subfloat[\label{}]{\includegraphics[width=4.0cm,height=3.0cm]{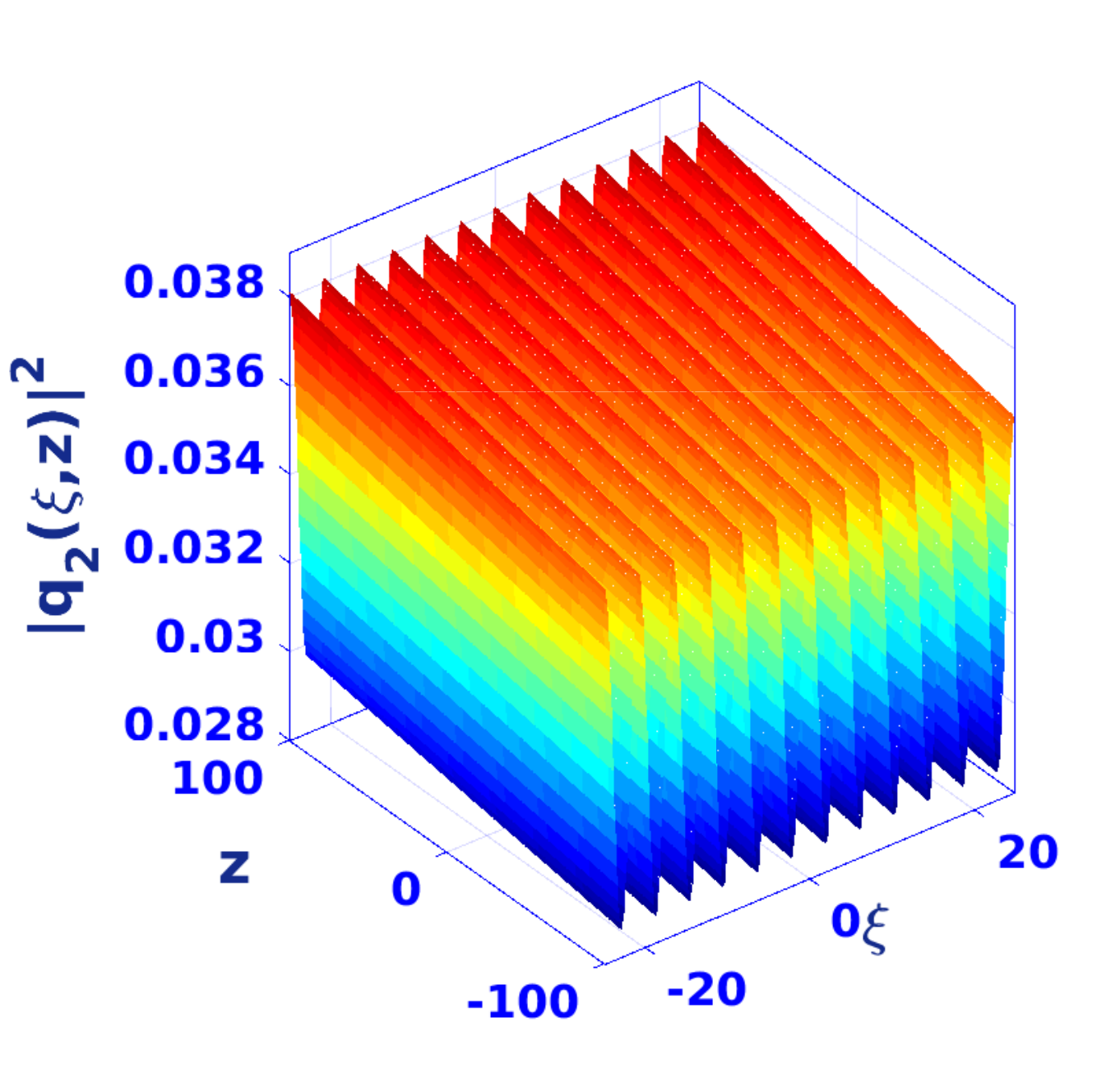}}
	\caption {For Solution I (Case(b)) for $m = 0.7$ (a) Plot of $|q_1|^2~\rm(blue~solid~line)$,~$|q_2|^2~\rm(red~dotted~line)$ versus $\xi$ for  $\beta=0.3,~\kappa=0.8,~v=1.5,~k_1=0.05,~k_2=0.01,~a_5=0.5,~c_1=-0.0143,~c_2=0.0133,~c_3=0.0702,~D=0.0183$ at $z=5$, (b) Plot of kinematic (black dashed line), higher order (red dotted line) and combined chirping (blue solid line) of $q_2$   versus $\xi$  for $\kappa=0.9,~\beta=0.4,~a_5=0.5,~v=3.5,~k_1=0.5,~k_2=0.4,~c_1=-0.00320559,~c_2=-0.00114485,~c_3=-0.276052,~D=-22.5694$, (c),(d)  Plots of  $\delta \omega_{1,2}$ versus $\kappa$ for $a_5=0.3$, $\delta \omega_{1,2}$ versus $a_5$ for $\kappa=0.5$ respectively and others parameters are $\beta=0.2,~v=0.9,~k_1=0.3,~k_2=0.1,~\xi=5$, 
	(e) Plot of  $|q_1|^2~\rm(black~dashed~line)$,~$|q_2|^2~\rm(red~dotted~line)$ versus $\kappa$ for  $\beta=0.3,~v=0.9,~k_1=0.1,~k_2=0.05,~a_5=0.5,~\xi=5$, (f),(g) Simulation of the intensity profiles for same parameter values as in (a).}
\end{figure}

{\bf Solution II.}\\ 
Another solution to Eqn.(\ref{11}) is given by
\begin{equation}\label{12}
\mu(\xi) = A^2 [1-m D sn^2(\xi, m)]\,,
\end{equation}
 provided
\begin{eqnarray}\label{13}
&&d_1 = (1+m) - \frac{3}{D},~~\eta_1A^2
=\frac{2}{D},\, \nonumber \\
&&\frac{2c_3}{A^2}= 4(1+m) - \frac{6}{D} - 2mD.
\end{eqnarray}
Fig.3(a) depicts the intensity profile of both the components with respect to $\xi$ while 
chirping reversal for $q_1$ is shown in Fig.3(b). Chirping of both the components increase as $\kappa$ increases as seen in Fig.3(c). Transition from negative to positive chirping is seen in the first component while the chirping of the second component remains negative as $\kappa$ increases. Fig.3(d) shows that the chirping of the first component decreases but that of the second component increases as $a_5$ increases. Transition from positive to negative chirping is seen in the first component but chirping of the second component remains negative as $a_5$ increases.
Both $|q_1|^2$ and $|q_2|^2$ decrease as $\kappa$ increases as can be seen from Fig.3(e). Simulation of the intensity profiles of both the components are shown in Figs.3(f) and (g) which clearly show the stability of the periodic wave with associated pulse compression.
\begin{figure}[]%fig3
	\centering
	\subfloat[\label{}]{\includegraphics[width=4.0cm,height=3.0cm]{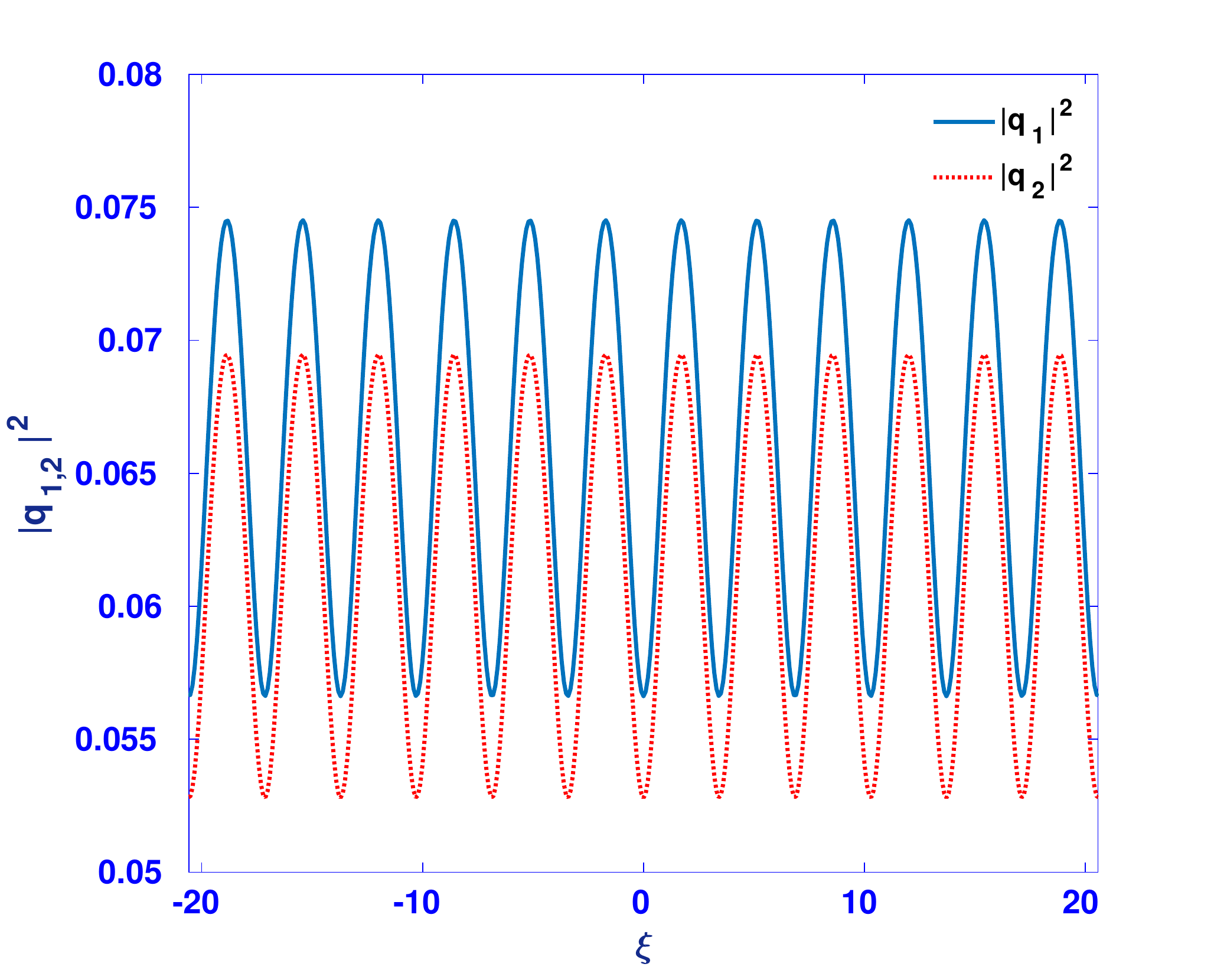}}
	~
	\subfloat[\label{}]{\includegraphics[width=4.0cm,height=3.0cm]{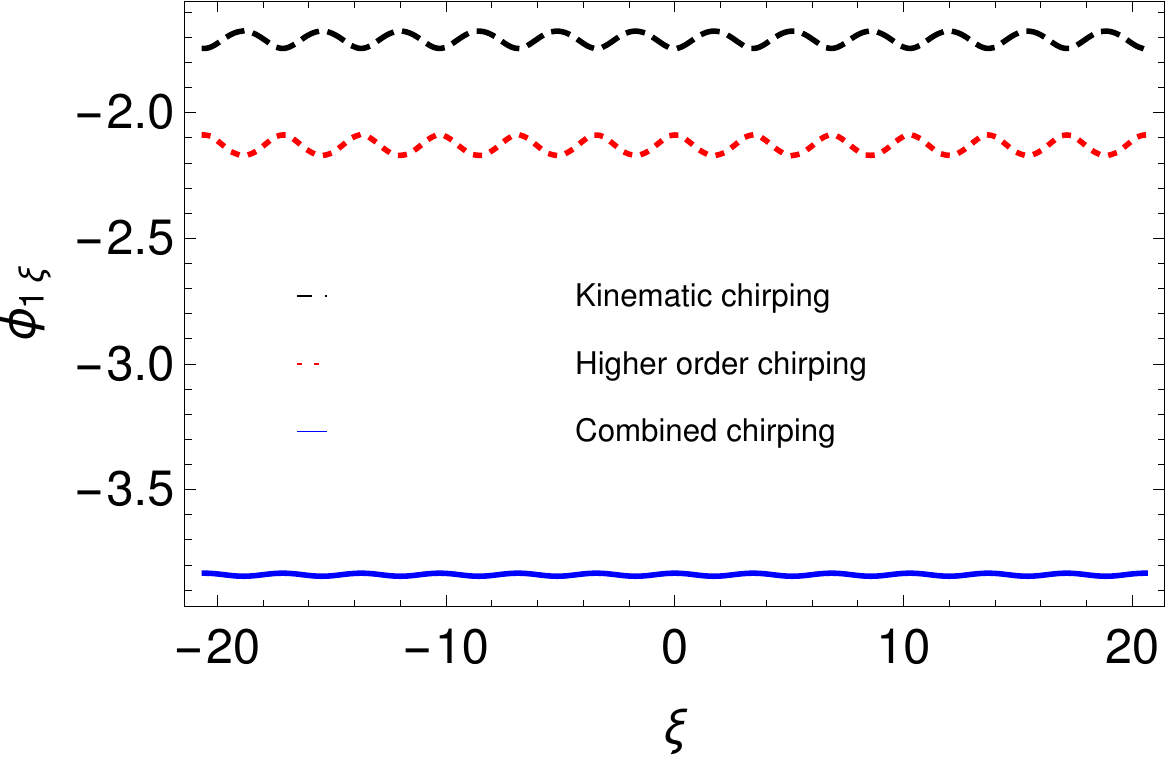}}
	~
	\subfloat[\label{}]{\includegraphics[width=4.0cm,height=3.0cm]{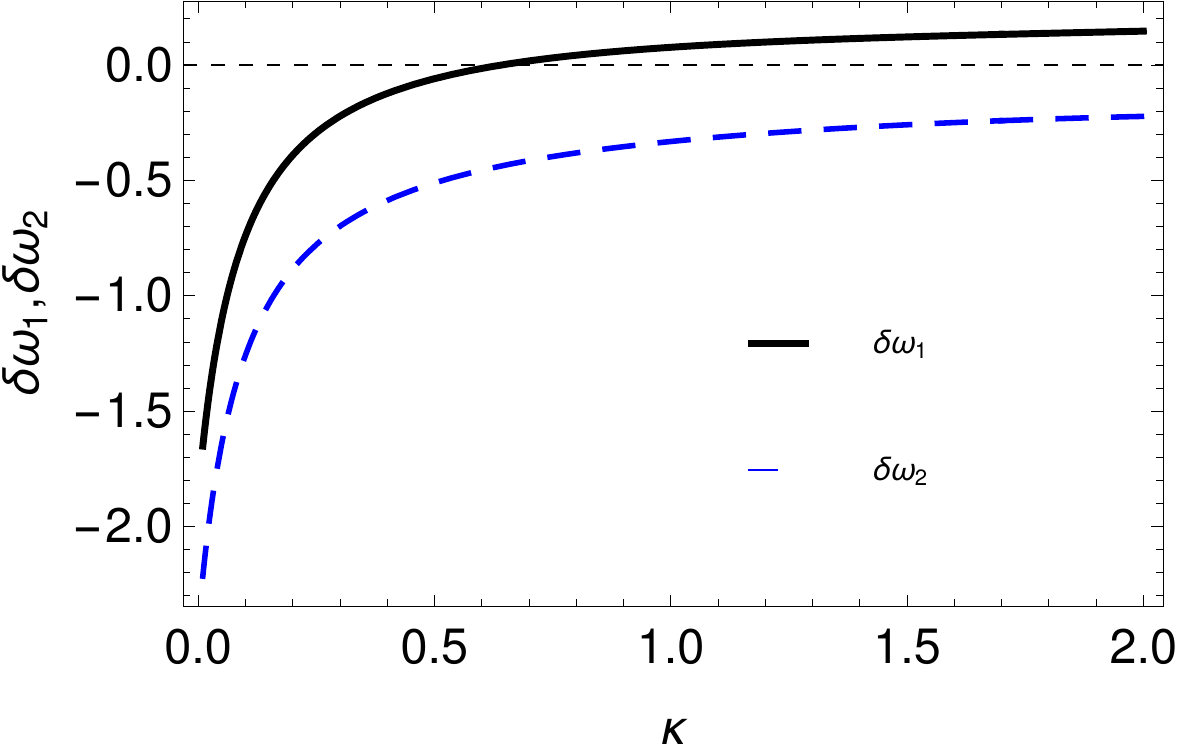}}
	~
	\subfloat[\label{}]{\includegraphics[width=4.0cm,height=3.0cm]{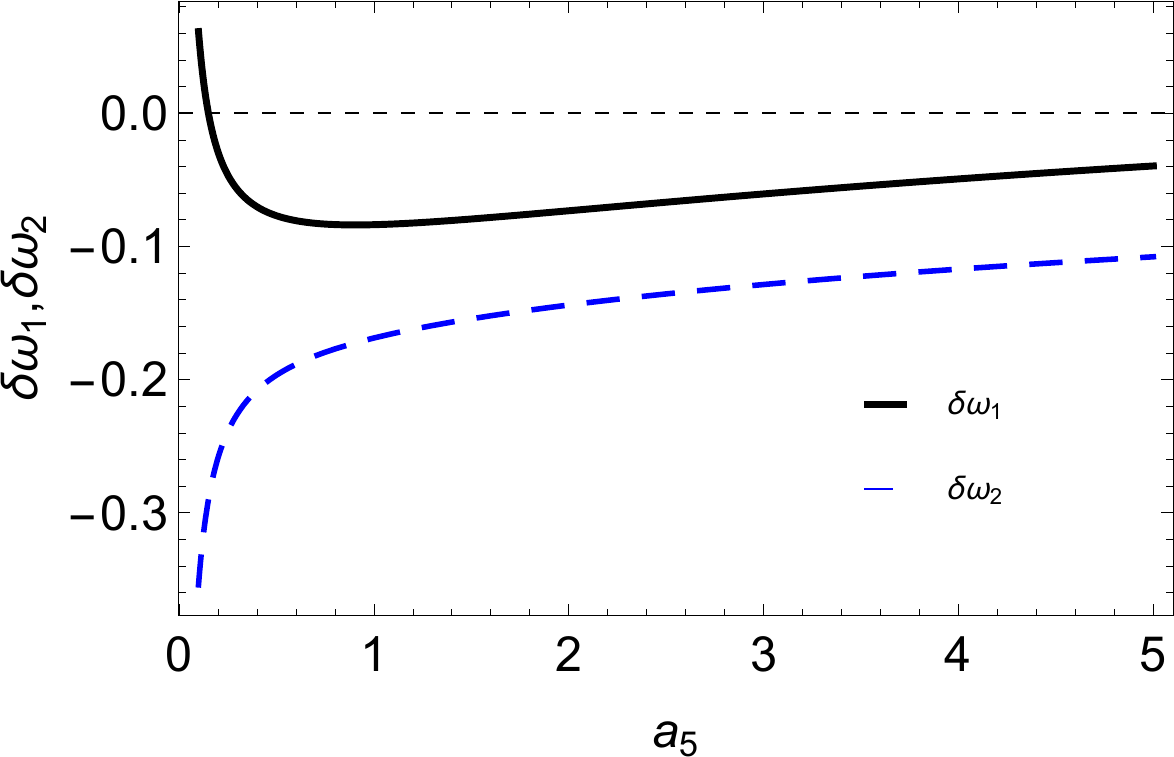}}\\
	\subfloat[\label{}]{\includegraphics[width=4.0cm,height=3.0cm]{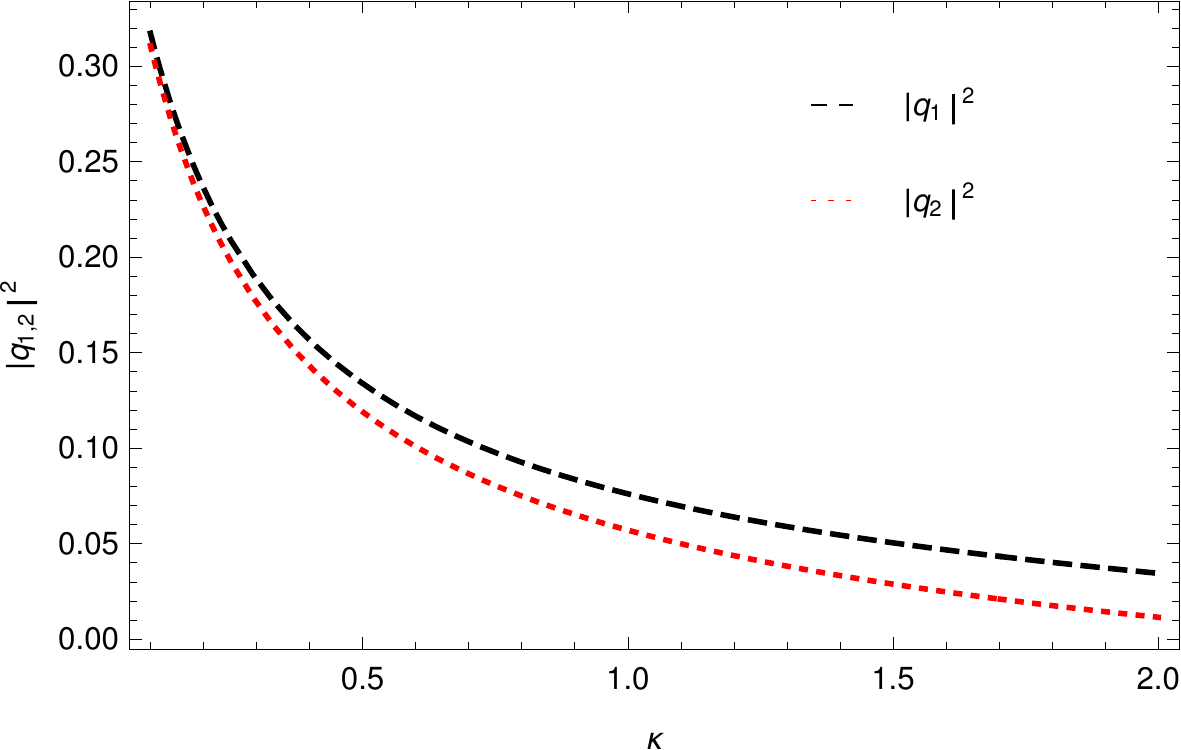}}
	\subfloat[\label{}]{\includegraphics[width=4.0cm,height=3.0cm]{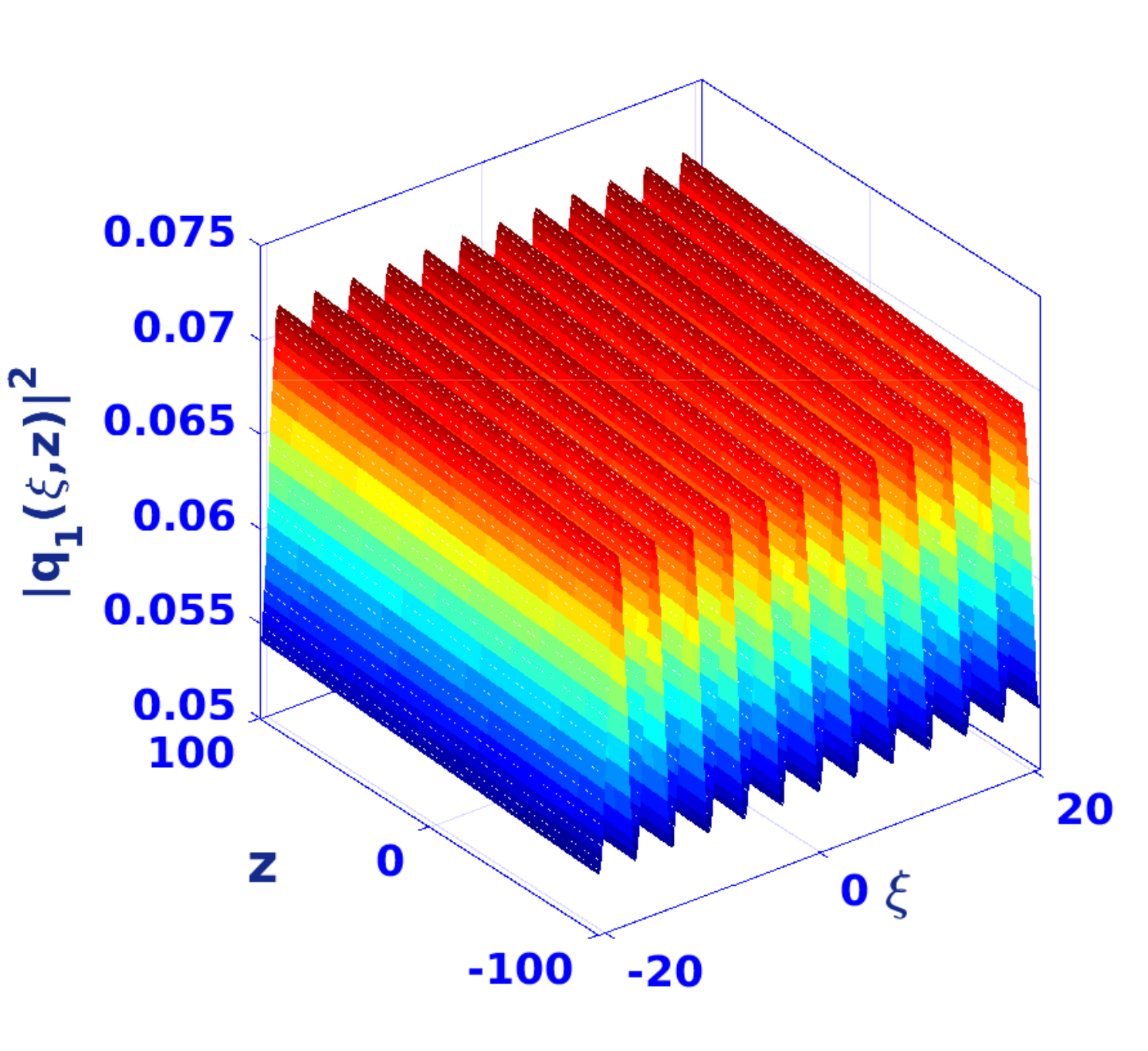}}
	~~
	\subfloat[\label{}]{\includegraphics[width=2.0cm,height=3.0cm]{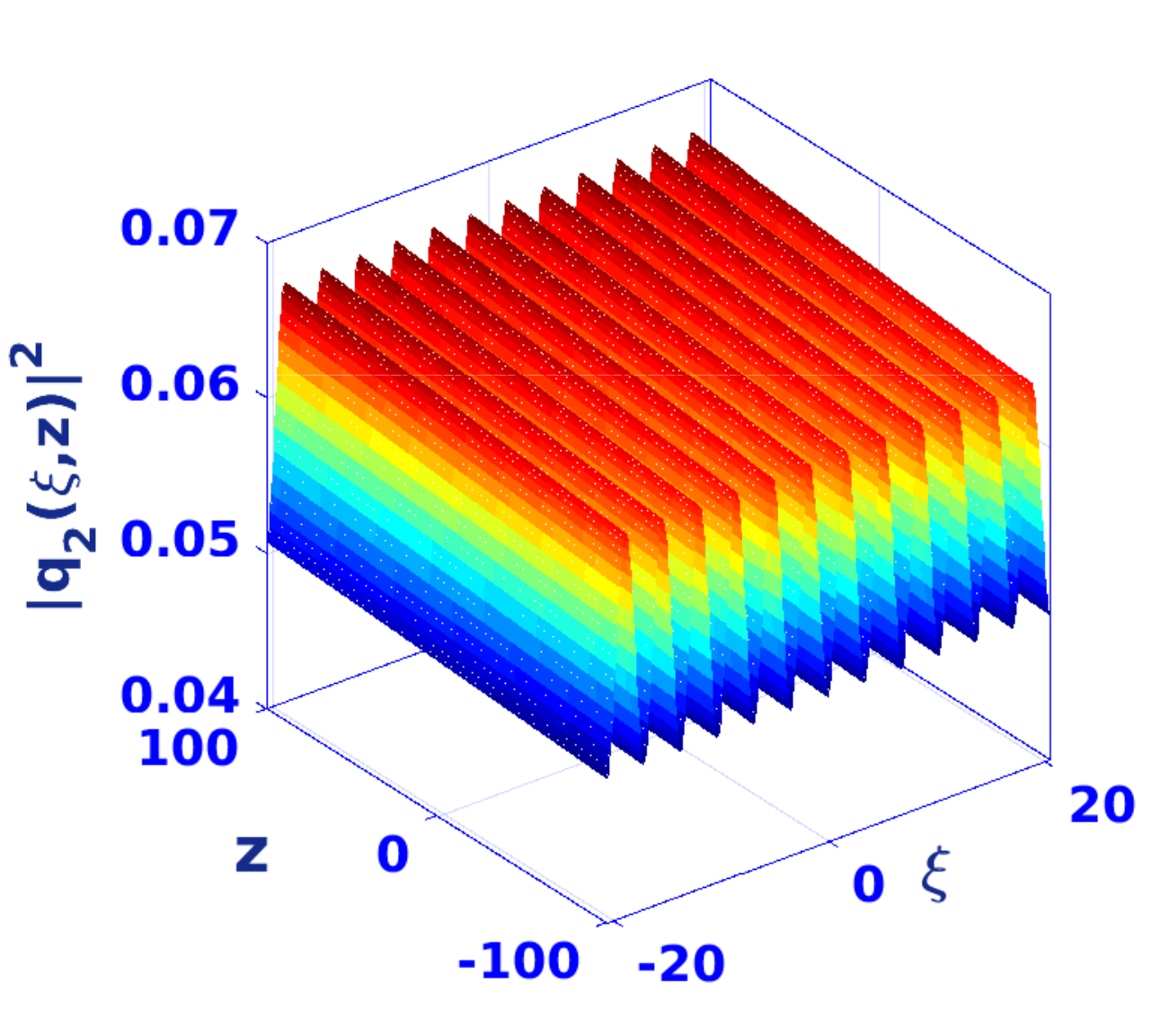}}
	\caption{Solution II for $m = 0.3$ (a) Plot of $|q_1|^2~\rm(blue~solid~line)$,~$|q_2|^2~\rm(red~dotted~line)$ versus $\xi$ for  $,~\beta=0.1,~\kappa=0.9,~v=2.5,~k_1=0.05,~k_2=0.01,~a_5=0.5,~c_1=-0.016,~c_2=0.0149,c_3=0.3261,~D=-1.0336$ at $z=5$, (b) Plot of kinematic chirping (black dashed line), higher order chirping (red dotted line) and combined chirping (blue solid line) of $q_1$ versus $\xi$  for $\kappa=0.6,~\beta=0.15,~a_5=0.5,~v=0.5,~k_1=0.8,~k_2=0.5,~c_1=-0.376088,~c_2=-0.0376068,~c_3=5.2821,~D=-0.128767$, (c),(d)  Plots of  $\delta \omega_{1,2}$ versus $\kappa$ for $a_5=0.1$, $\delta \omega_{1,2}$ versus $a_5$ for $\kappa=0.9$ respectively and other parameters are $\beta=0.1,~v=2.5,~k_1=0.2,~k_2=0.1,~\xi=5$,  
	(e) Plot of  $|q_1|^2~\rm(black~dashed~line)$,~$|q_2|^2~\rm(red~dotted~line)$ versus $\kappa$ for  $\beta=0.1,~v=2.5,~k_1=0.2,~k_2=0.1,~a_5=0.5,~\xi=5$, (f),(g) Simulation of the intensity profiles for same parameter values as in (a)}.	
\end{figure}

{ \bf Solution III.}\\
Yet another solution to Eqn.(\ref{11}) is
\begin{equation}\label{16}
\mu(\xi) = \frac{A^2}{[1+B dn(\xi, m)]}\,,
\end{equation}
provided either
$B=1,~\text{or~}B = (1-m)^{-1/2}$.\\
{\bf Case (a). $B = 1$} \\
The constraint relations in this case are given by
\begin{equation}\label{17}
d_1=\frac{4 + m}{4},~~~\eta_1 A^2 = -m,~~~
\frac{c_3}{A^2}=1\,.
\end{equation}
The intensity profiles of both the components and chirp reversal for the component $q_1$ with respect to $\xi$ are shown in Figs.4(a) and (b) respectively. The chirping of the first component $\delta\omega_1$ initially increases with increasing $\kappa$ but starts decreasing after a certain value of $\kappa$ is reached while that of the second component $\delta\omega_2$ increases as $\kappa$ increases (Fig.4(c)). $\delta\omega_1$ ($\delta\omega_2$) shows transition from positive (negative) to negative (positive) chirping as $\kappa$ increases. The later behaviour of $\delta\omega_1$ and $\delta\omega_2$ is also seen when $a_5$ increases (Fig.4(d)). Fig.4(e) shows that intensity of both the components decrease as $\kappa$ increases. Fig.4(f) and (g) demonstrate stable evolution with pulse compression.\\ 
\begin{figure}
	\centering
	\subfloat[\label{}]{\includegraphics[width=4cm,height=3.0cm]{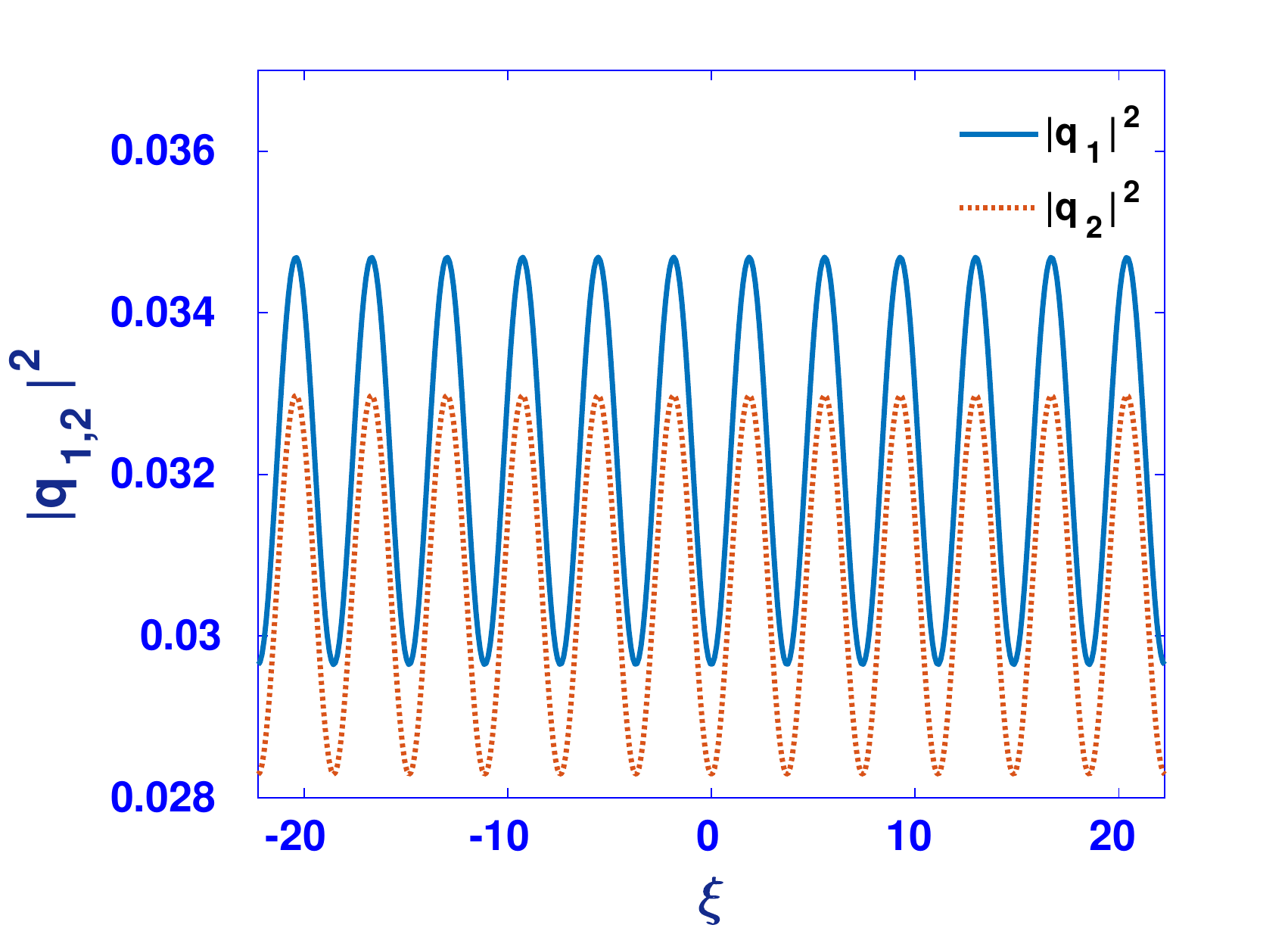}}
	~
	\subfloat[\label{}]{\includegraphics[width=4cm,height=3.0cm]{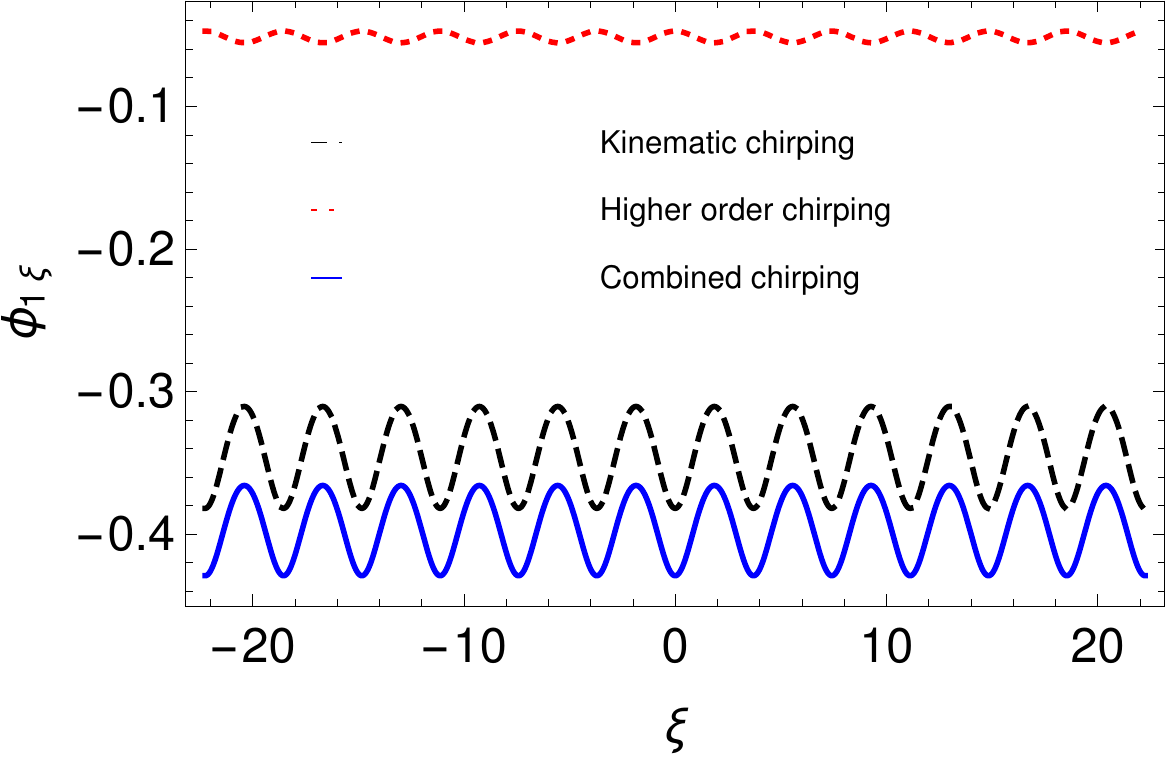}}
	~
	\subfloat[\label{}]{\includegraphics[width=4cm,height=3.0cm]{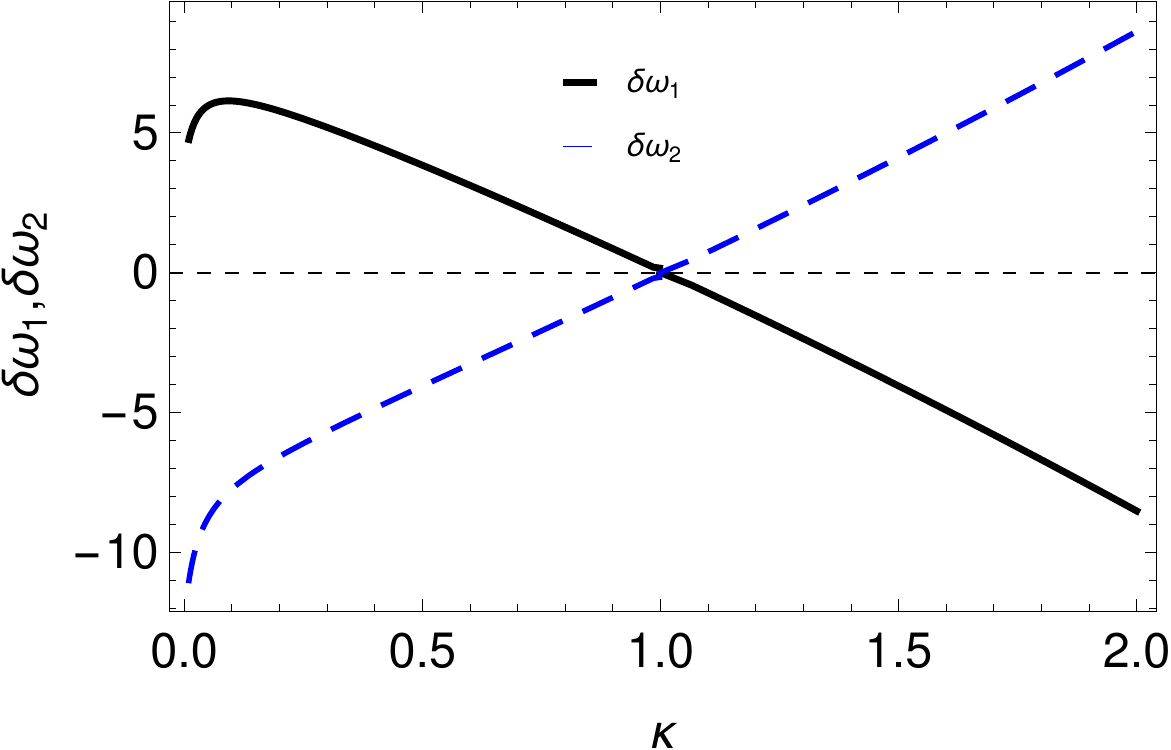}}
	~
	\subfloat[\label{}]{\includegraphics[width=4cm,height=3.0cm]{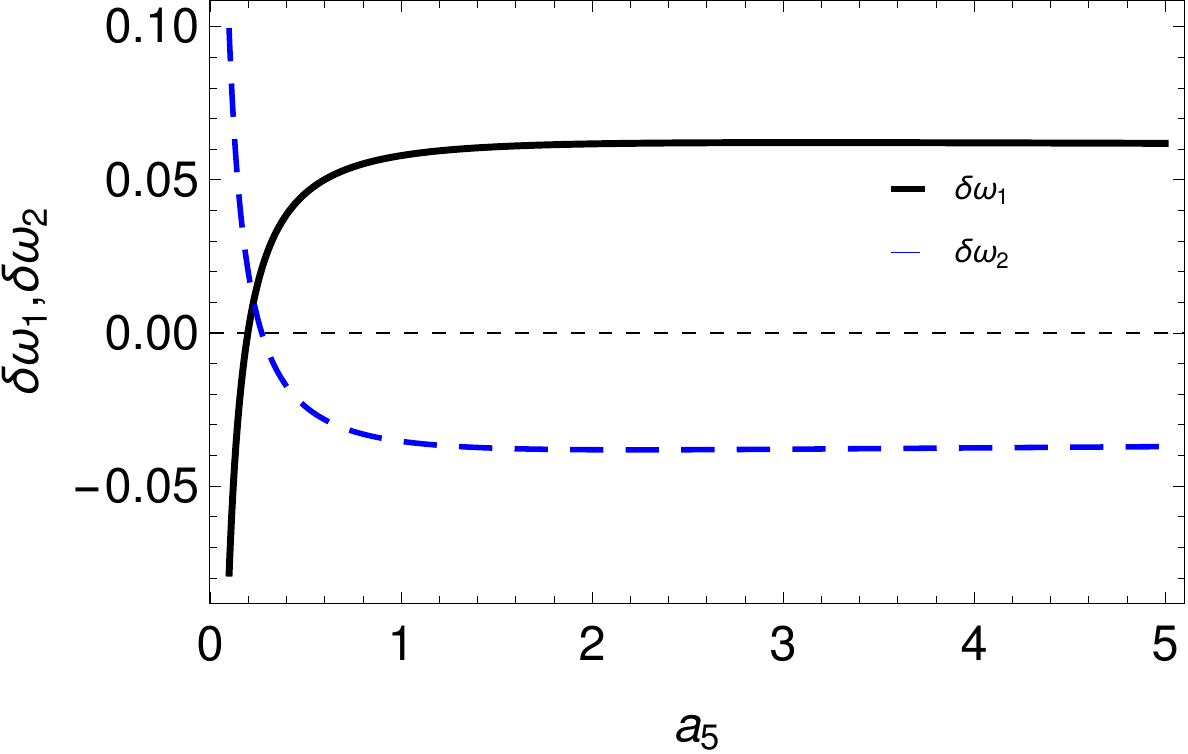}}\\
	\subfloat[\label{}]{\includegraphics[width=4cm,height=3.0cm]{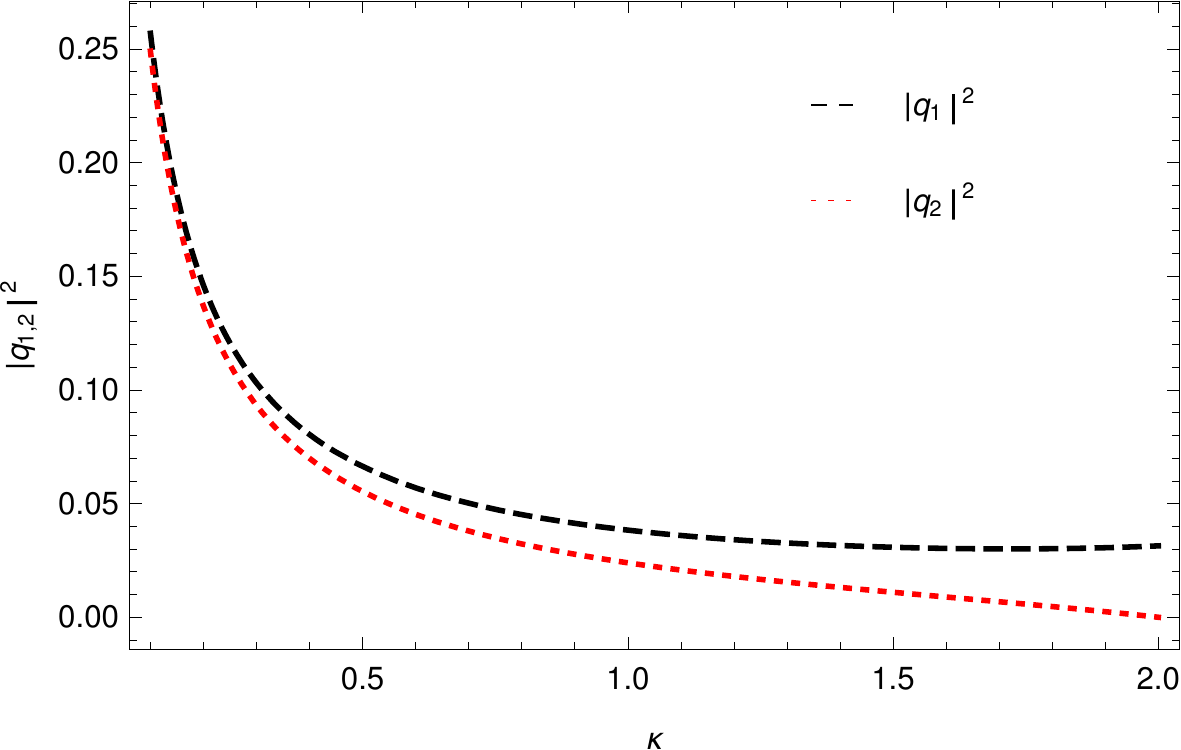}}
	~~
	\subfloat[\label{}]{\includegraphics[width=4.0cm,height=3.0cm]{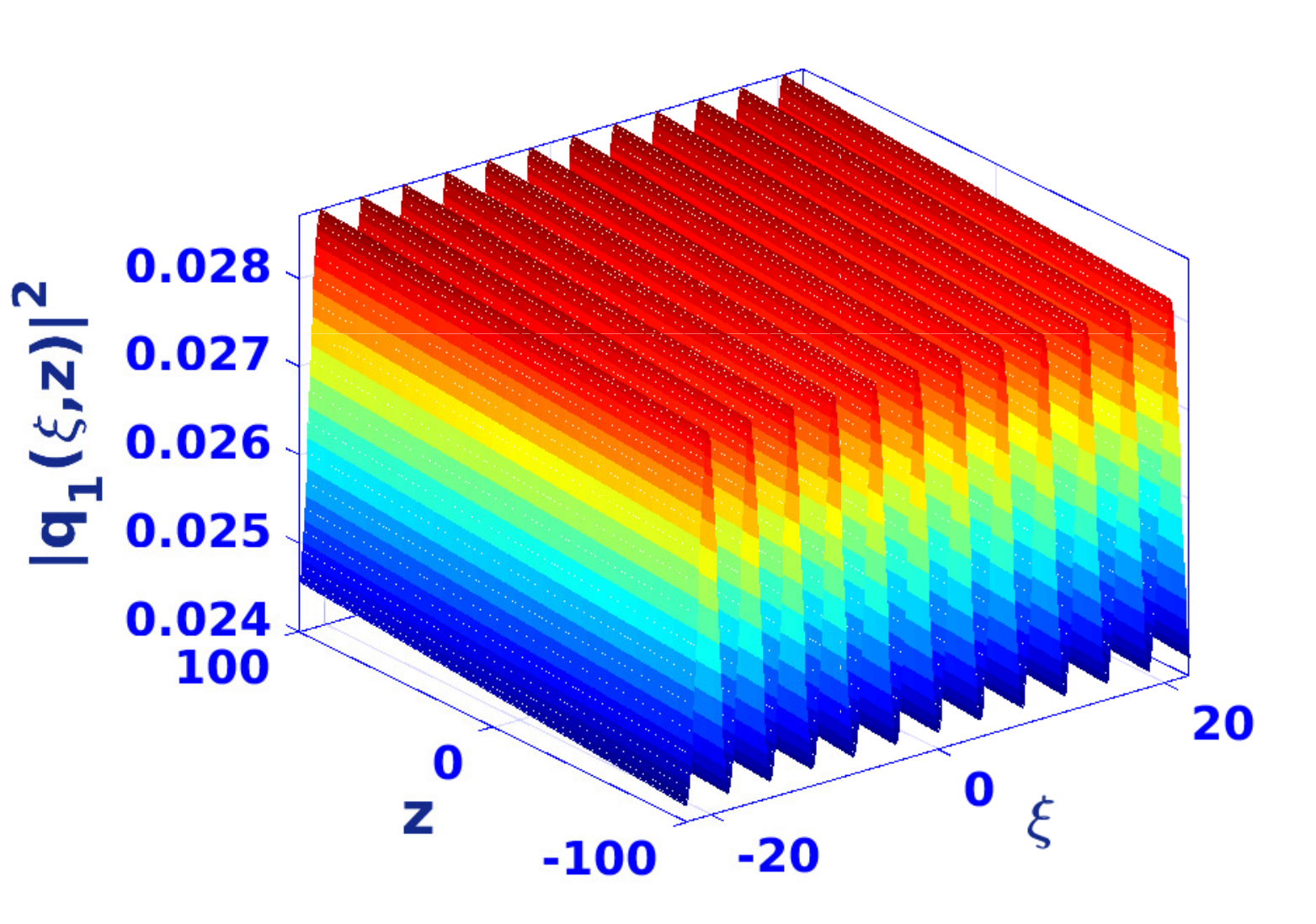}}
	~~
	\subfloat[\label{}]{\includegraphics[width=4.0cm,height=3.0cm]{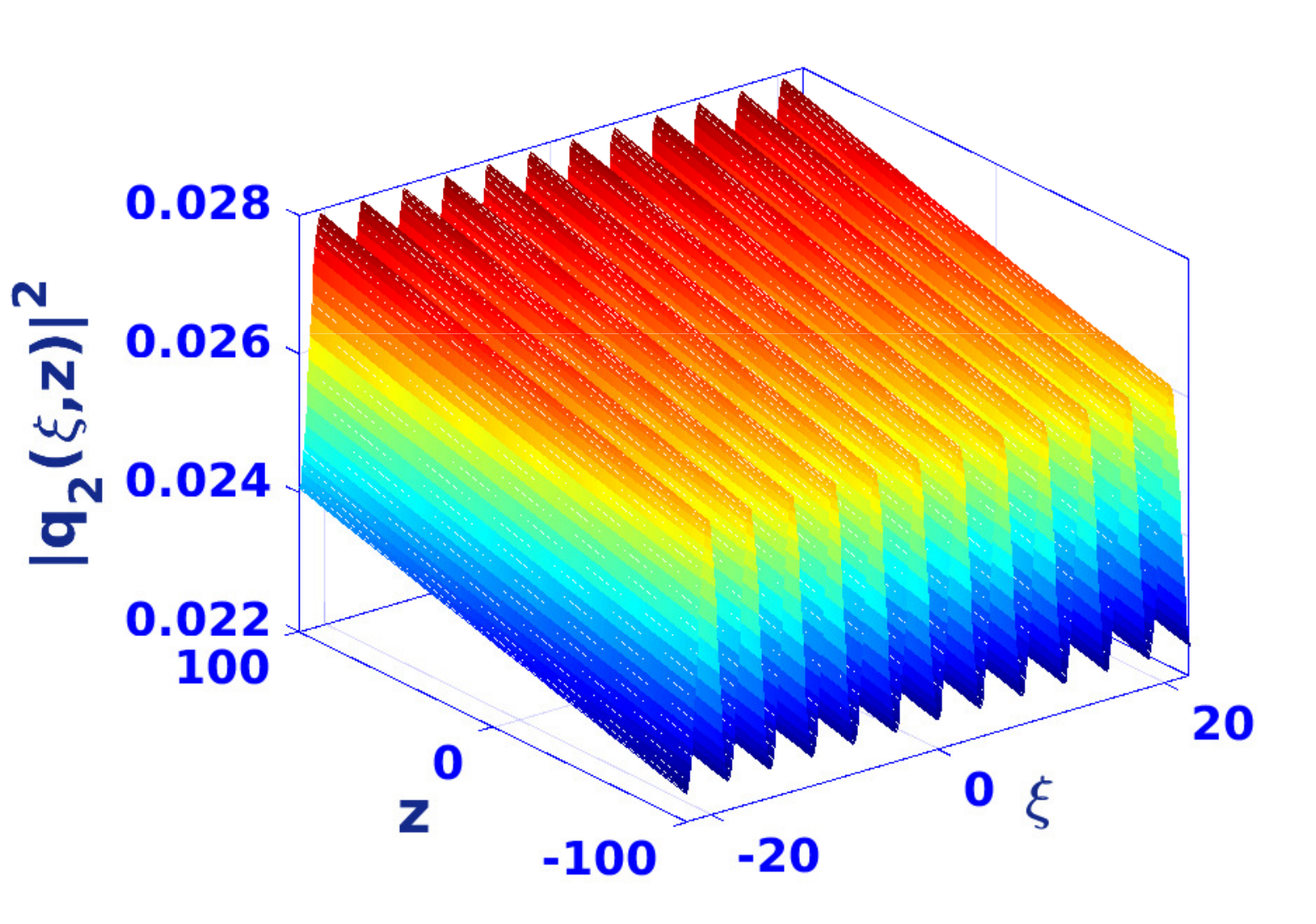}}
	\caption{For Solution III (Case (a)) for $m=0.5$, (a) Plot of $|q_1|^2~\rm(blue~solid~line)$,~$|q_2|^2~\rm(red~dotted~line)$ versus $\xi$ for  $\kappa=0.9,~v=2.5,~\beta=0.1933,~k_1=0.05,~k_2=0.01,~a_5=0.05,~c_1=-0.08,~c_2=0.0769,~c_3=0.0588$ at $z=5$, (b) Plot of kinematic chirping (black dashed line), higher order chirping (red dotted line) and combined chirping (blue solid line) of $q_2$ versus $\xi$  for $\kappa=0.9,~a_5=0.95,~v=4.5,~k_1=0.5,~k_2=0.1,~c_1=-0.0258548,~c_2=-0.00315303,~c_3=0.106114$, (c),(d)  Plots of  $\delta \omega_{1,2}$ versus $\kappa$ for $a_5=0.02$, $\delta \omega_{1,2}$ versus $a_5$ for $\kappa=1.1$ respectively and others parameters are $v=4.5,~k_1=0.9,~k_2=0.1,~\xi=5$,   
		(e) Plot of  $|q_1|^2~\rm(black~dashed~line)$,~$|q_2|^2~\rm(red~dotted~line)$ versus $\kappa$ for  $v=4.5,~k_1=0.25,~k_2=0.1,~a_5=0.02,~\xi=5$, (d),(e) Simulation of the intensity profiles for same parameter values as in (a).} 
	
\end{figure}
{\bf Case (b). $B = (1-m)^{-1/2}$}\\
The parametric conditions to be satisfied are the following
\begin{equation}\label{18}
d_1=\frac{4-5m}{4},~~~\eta_1A^2 = m,~~~
\frac{c_3}{A^2}= (1-m)\,.
\end{equation}
The intensity profile of both the components with respect to $\xi$ is shown in Fig.5(a).
Chirp reversal for the component $q_2$ is shown in Fig.5(b). $\delta\omega_1$ increases as $\kappa$ increases and there is transition from negative to positive chirping as is seen in Fig.5(c). $\delta\omega_2$ initially increases with increasing $\kappa$ but starts decreasing when a certain value of $\kappa$ is reached but unlike $\delta\omega_1$, it remains negative as $\kappa$ increases. Fig.5(d) shows that $\delta\omega_1$ decreases as $a_5$ increases and there is transition from positive to negative chirping for increasing $a_5$. But $\delta\omega_2$ increases and remains negative as $a_5$ increases.
The intensity of both the components decreases as $\kappa$ increases (Fig.5(e)). Figs.5(f) and (g) display pulse compression as well as stable evolution of the periodic wave for the chosen values of the parameters.  
\begin{figure}
	\centering
	\subfloat[\label{}]{\includegraphics[width=4.0cm,height=3.0cm]{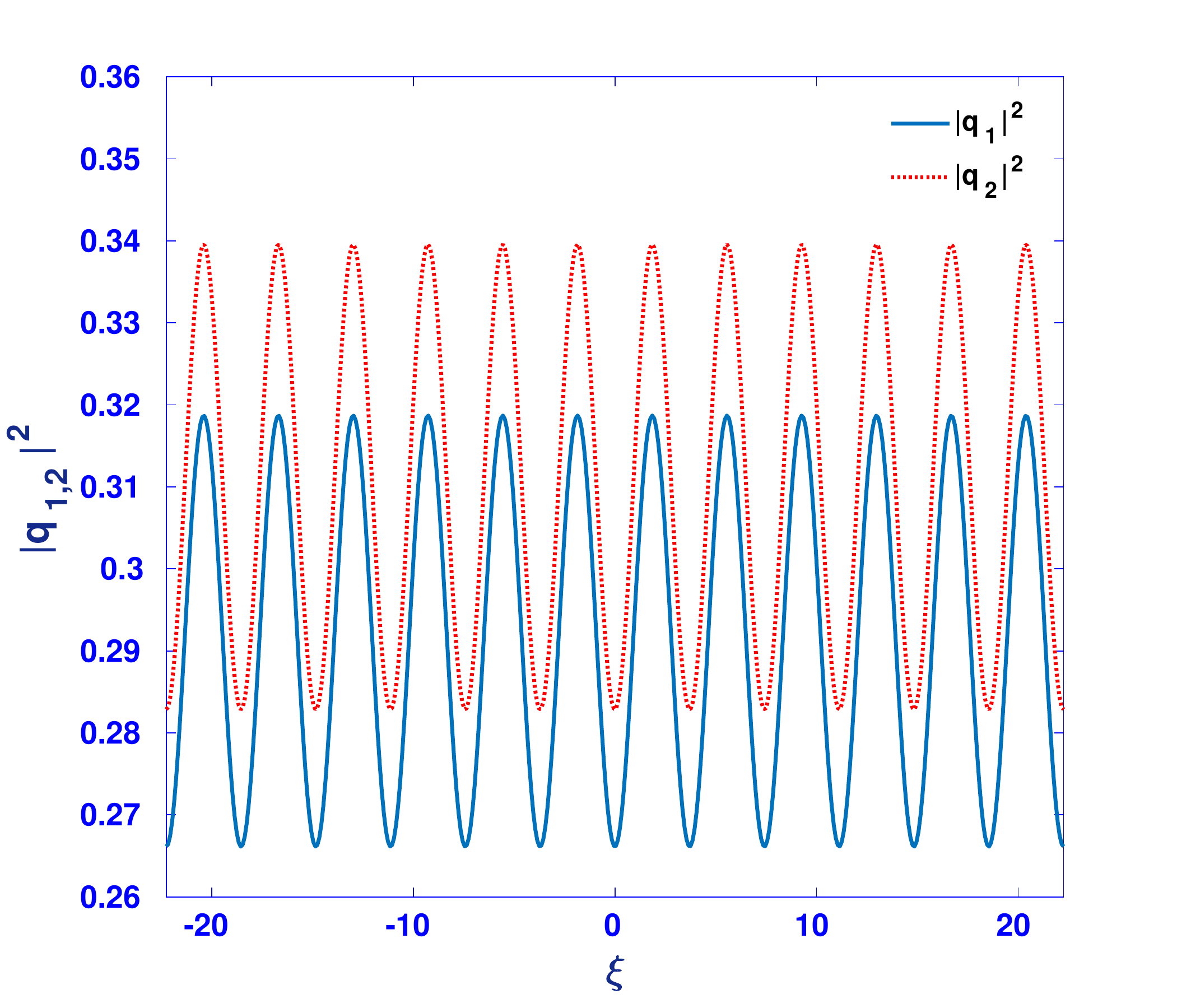}}
	~
	\subfloat[\label{}]{\includegraphics[width=4.0cm,height=3.0cm]{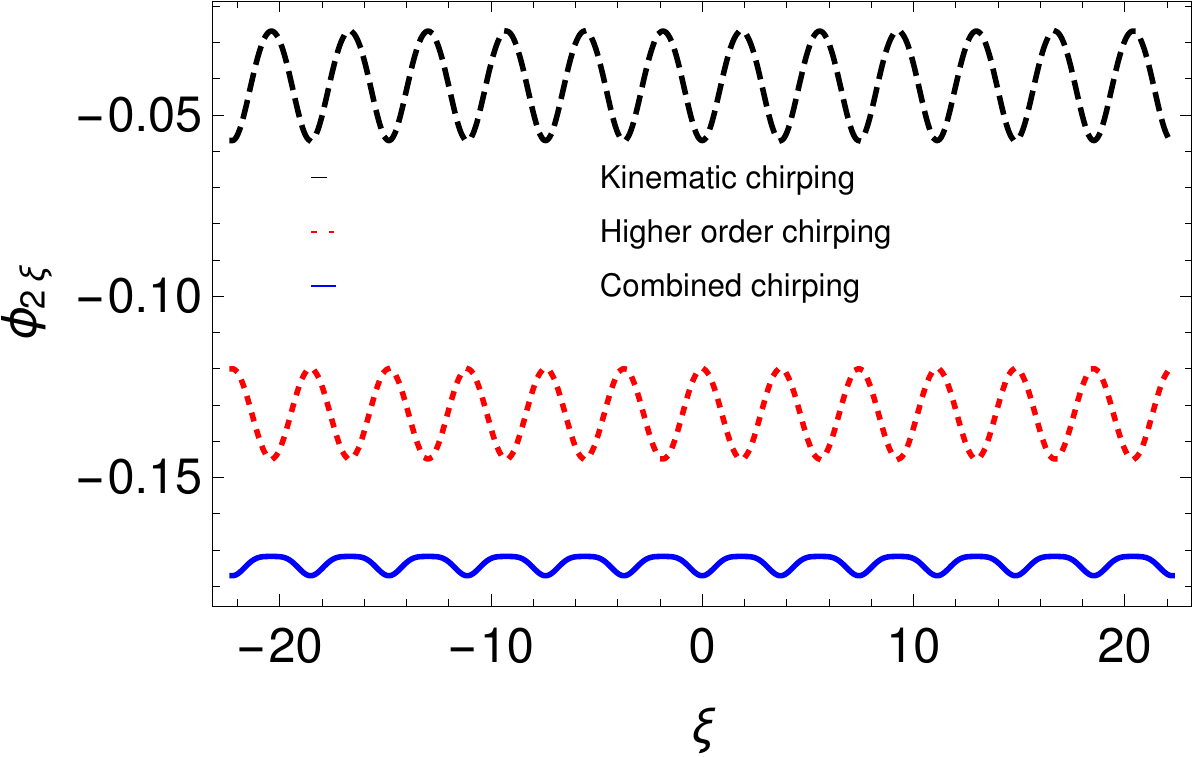}}
	~
	\subfloat[\label{}]{\includegraphics[width=4.0cm,height=3.0cm]{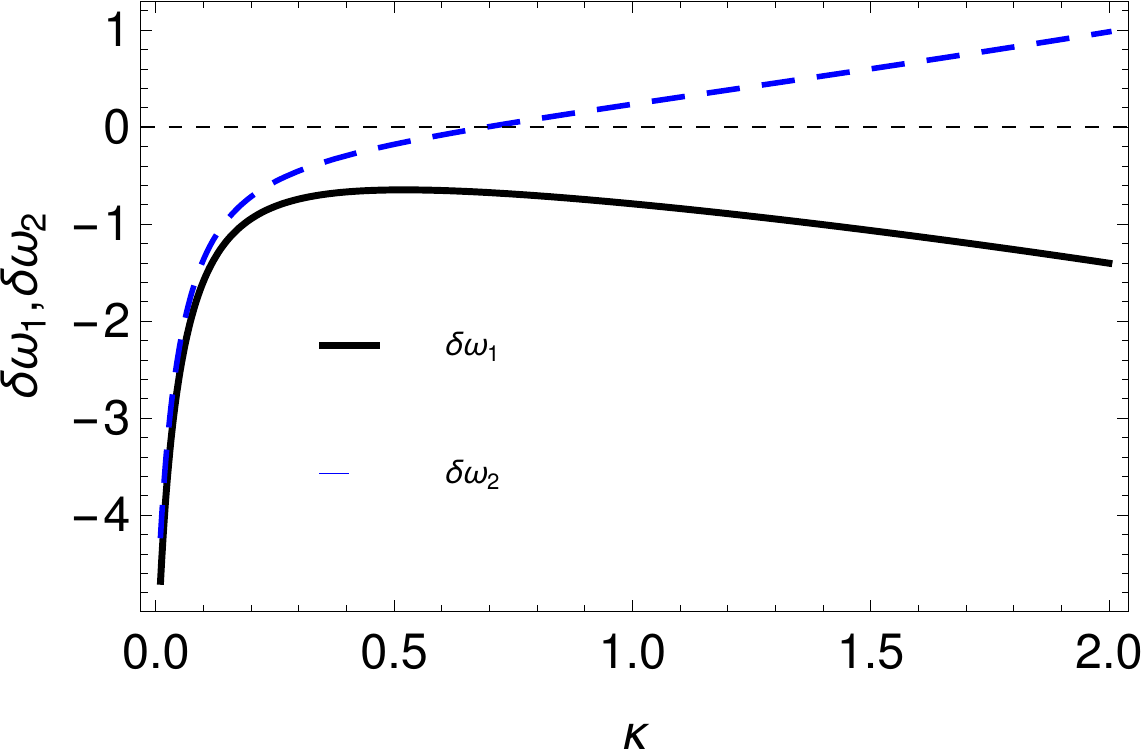}}
	~
	\subfloat[\label{}]{\includegraphics[width=4.0cm,height=3.0cm]{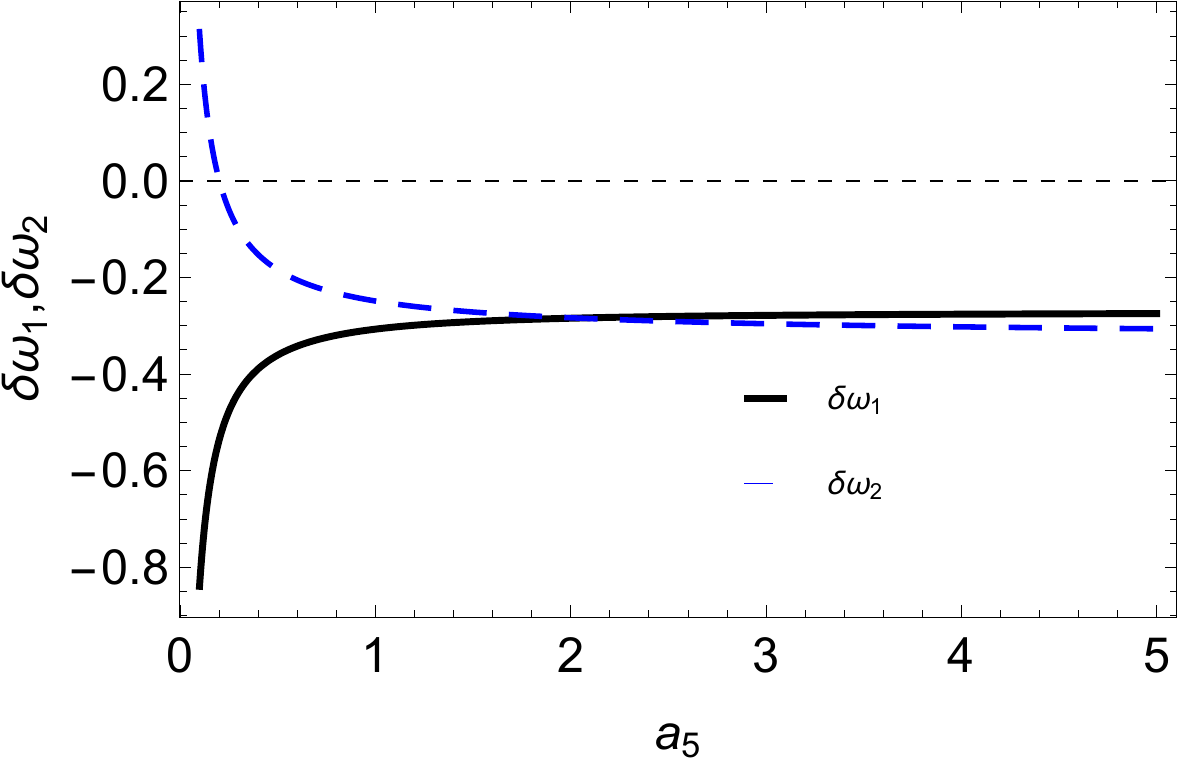}}\\
	\subfloat[\label{}]{\includegraphics[width=4.0cm,height=3.0cm]{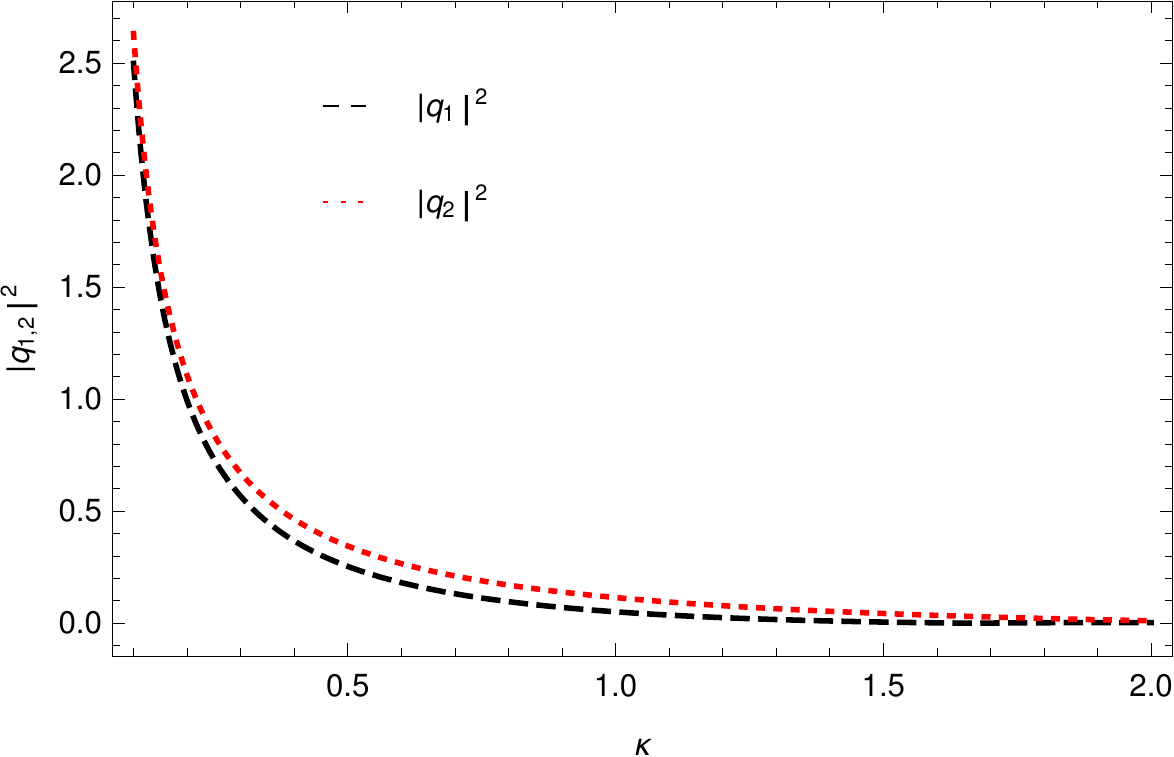}}
	~~
	\subfloat[\label{}]{\includegraphics[width=4.0cm,height=3.0cm]{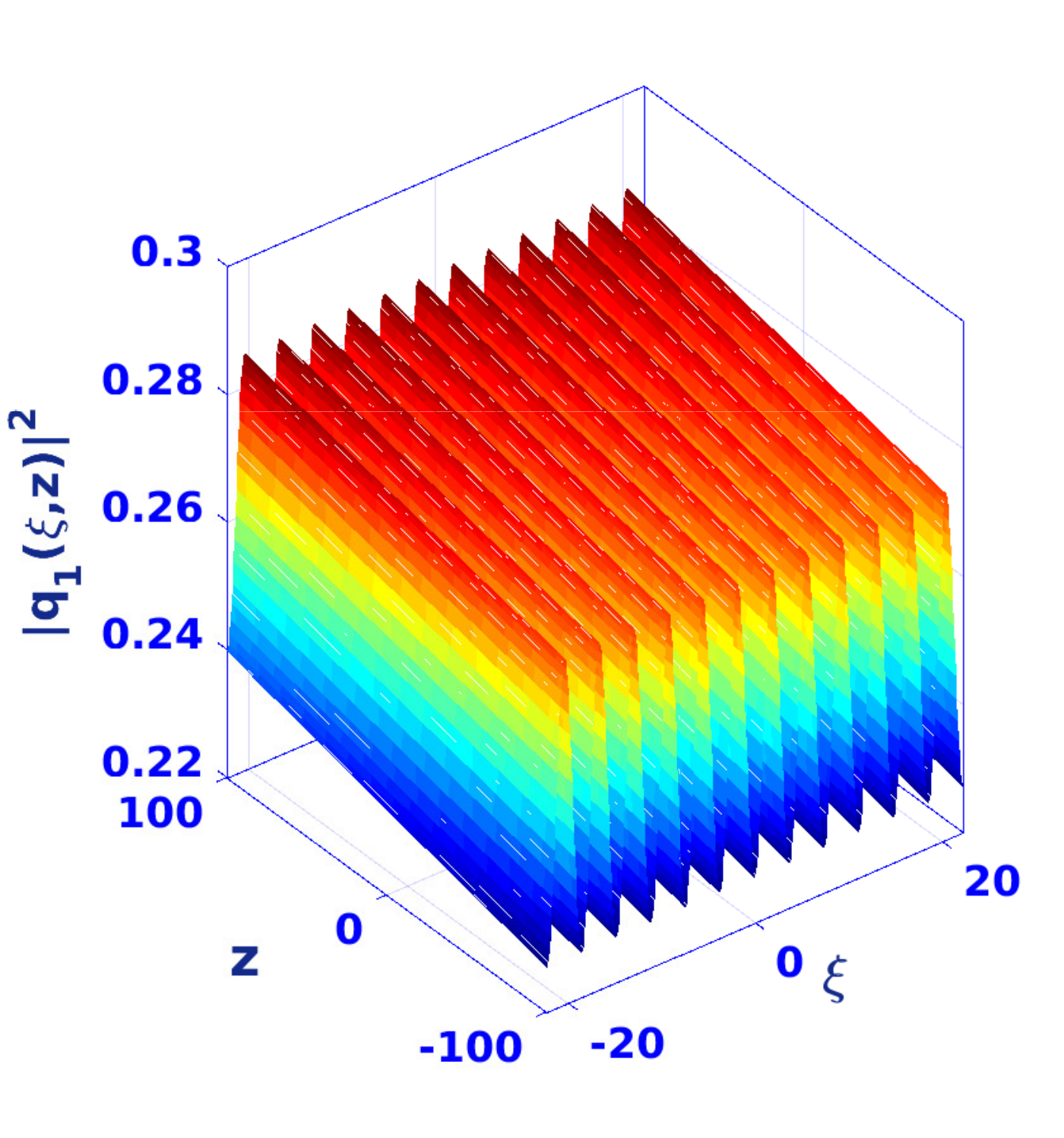}}
	~~
	\subfloat[\label{}]{\includegraphics[width=4.0cm,height=3.0cm]{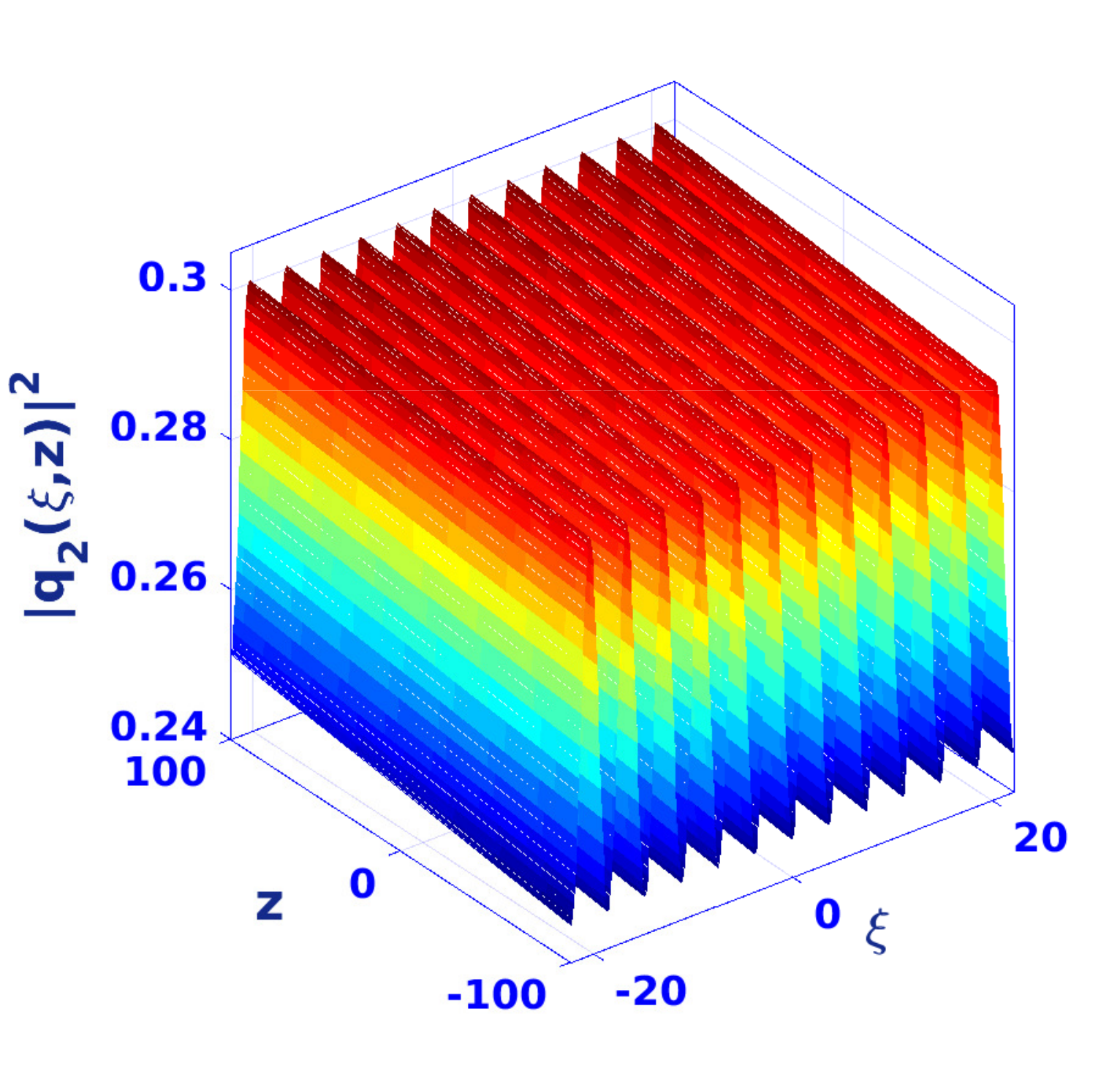}}
	\caption{For Solution III (Case(b)) for $m=0.5$, (a) Plot of $|q_1|^2~\rm(blue~solid~line)$,~$|q_2|^2~\rm(red~dotted~line)$ versus $\xi$ for  $\kappa=0.9,~v=0.1,~\beta=0.4128,~k_1=0.01,~k_2=0.05,~a_5=0.05,~c_1=0.4331,~c_2=-0.4674,~c_3=0.3121$ at $z=5$, (b) Plot of kinematic chirping (black dashed line), higher order chirping (red dotted line) and combined chirping (blue solid line) of $q_2$ versus $\xi$  for $\kappa=0.8,~a_5=0.5,~v=2.5,~k_1=0.2,~k_2=0.5,~c_1=-0.0130647,~c_2=-0.0444198,~c_3=0.0896896$, (c),(d)  Plots of  $\delta \omega_{1,2}$ versus $\kappa$ for $a_5=0.1$, $\delta \omega_{1,2}$ versus $a_5$ for $\kappa=1.1$ respectively and others parameters are $v=3.5,~k_1=-0.2,~k_2=-0.8,~\xi=5$,   
	(e) Plot of  $|q_1|^2~\rm(black~dashed~line)$,~$|q_2|^2~\rm(red~dotted~line)$ versus $\kappa$ for  $v=0.05,~k_1=0.05,~k_2=0.3,~a_5=0.1,~\xi=5$, (f),(g) Simulation of the intensity profiles for same parameter values as in (a).} 
	
\end{figure}
\section{Conclusion}
In this paper we have considered 
the nonlinear coupled cubic Helmholtz system in the presence of the non-Kerr 
terms like self steepening and self frequency shift. This 
system describes nonparaxial pulse propagation with Kerr and non-Kerr 
nonlinearities with spatial dispersion originating from the nonparaxial effect 
that comes into play when the slowly varying envelope approximation fails. We 
have obtained three families of exact chirped elliptic wave solutions.
The parametric conditions for the existence of these solutions are 
given. 
The novel physical consequence of the presence of the non-Kerr nonlinearities 
occurs in the chirping property of the periodic waves. Specifically, for a 
particular relation between the self steepening and the self frequency shift 
parameters, these solutions are characterized by nontrivial phase and have two 
intensity dependent chirping terms. One term varies as the reciprocal of the 
intensity while the other term, which depends on the non-Kerr nonlinearities, 
is directly proportional to the intensity. As a result, chirp reversal occurs 
across the wave profile. For a different relation between the non-Kerr terms, 
there exists only one intensity dependent term which is inversely proportional 
to the intensity, resulting in chirping but no chirp reversal.
Study of the effect of nonparaxial and non-Kerr nonlinearity parameters on 
chirping of the obtained solutions reveals dissimilar character of chirping of 
the two compoinents. Specifically, the effect of the nonparaxial parameter 
gives rise to positive chirping in one of the components and negative chirping 
in the other component. This behaviour is also seen in some cases as the effect
of self frequency shift (self steepening) parameter. Also the variation of the 
nonparaxial parameter and non-Kerr nonlinearity sometimes result in transition 
from the positive (negative) to the negative (positive) chirping in one or 
both the components.
Effect of nonparaxial parameter on physical quantities like intensity, speed and pulse-width of the solitary wave solutions have been examined.
It is found that the speed of the solitary wave can be tuned by altering the 
nonparaxial parameter. 
The stability of the elliptic waves has been studied by means of direct simulations of the 
perturbed evolution of the solutions and are found to be stable for the parametric regions examined herein.  It is important to mention that we have obtained stable evolutions of the elliptic waves for appropriate choice of parameters even in case there is chirping but no chirp reversal. The nonlinearly chirped solutions 
presented here may find applications in nonparaxial self focusing of high power
laser beams in nonlinear media \cite{jin} and also where pre chirp managed 
femtosecond pulses are used for the pulse compression \cite{cun1} as well as for the generation of the widely tunable femtosecond 
pulses by pre chirp managed self phase modulation enabled spectral selection 
\cite{chang} in nonparaxial context.\\
As for future directions of study, it would be interesting to obtain 
chirped solutions of the coupled Helmholtz equation in the presence of the 
quintic, septic non-kerr nonlinearities. This is important in order to adapt to
the current progress in high repetition rate (beyond ultrashort, even 
autosecond) pulse sources based on the fiber technology \cite{hadrich}. Study 
of modulation instability in this context surely deserves investigation. Also, 
as identified by Peregrine \cite{peregrine}, modulation instability plays a key
role in the formation of patterns resembling freak waves or rogue waves 
\cite{chaudhuri}. Using physics-informed neural network (PINN) 
to obtain femtosecond nonparaxial optical solitons of the higher order 
nonlinear Helmholtz equation and to obtain fractional solitons in nonparaxial 
regime are other areas worth investigating as have been done for NLSE in 
\cite{fang} and \cite{wu} respectively. Some of these issues are currently 
under investigation and we hope to report on some of them in the near future.\\ \\
{\bf Data Availability Statement}
No new data were created or analysed in this study.

\end{document}